\documentclass[aps,graphicx,6pt,showkeys,showpacs,onecolumn]{revtex4}
\usepackage{amsmath}
\usepackage{amscd}
\usepackage{graphicx}
\usepackage{subfigure}
\usepackage{appendix}
\usepackage{amsfonts}
\usepackage{multirow}

\begin{document}



\title{Improving the stimulated Raman adiabatic passage via
dissipative quantum dynamics}

\author{Qi-Cheng Wu,$^{1}$ Ye-Hong Chen,$^{1}$ Bi-Hua Huang,$^{1}$ Jie Song,$^{2}$ Yan Xia,$^{1,*}$ Shi-Biao Zheng$^{1}$}

\address{$^{1}$Department of Physics, Fuzhou University, Fuzhou 350002, China\\
             $^{2}$Department of Physics, Harbin Institute of Technology, Harbin 150001, China}

\email{$^*$xia-208@163.com} 



\begin{abstract} We propose a method to improve the stimulated Raman adiabatic
passage (STIRAP) via dissipative quantum dynamics, taking into
account the dephasing effects. Fast and robust population transfer
can be obtained with the scheme by the designed pulses and detuning,
even though the initial state of the system is imperfect. With a
concrete three-level system as an example, the influences of the
imperfect initial state, variations in the control parameters, and
various dissipation effects are discussed in detail. The numerical
simulation shows that the scheme is insensitive to moderate
fluctuations of experimental parameters and the relatively large
dissipation effects of the excited state. Furthermore, the dominant
dissipative factors, namely, the dephasing effects of the ground
states and the imperfect initial state are no longer undesirable, in
fact, they are the important resources to the scheme. Therefore, the
scheme could provide more choices for the realization of the
complete population transfer in the strong dissipative fields where
the standard stimulated Raman adiabatic passage or shortcut schemes
are invalid.\end{abstract}


\maketitle


\section{Introduction}

The ability to accurately control a quantum system is a
fundamental requirement in many areas of modern science ranging
from quantum information processing~\cite{Barz} and coherent
manipulation of quantum systems~\cite{Scully1997} to
high-precision measurements~\cite{Kasevich2002,Kotru2015}. The
technique of stimulated Raman adiabatic passage (STIRAP) offers,
in principle, a simple approach for a complete population transfer
of a three-level quantum system has already drawn great attention,
and the basic properties of the STIRAP have been scrutinized both
theoretically and
experimentally~\cite{firstSTIRAP,Bergmann1998,Vitanov2001,Demirplak2003,Giannelli2014,Sun2014,Demirplak2002,Ivanov2004,Lacour2008,Lu2013,Issoufa2014,paraoanunc2016,Graefe2013}.
However, this process requires large pulse areas, i.e., large
average Rabi frequencies, and long interaction time. For several
applications, this requirement is a critical disadvantage, more
specifically, the ideal robustness and the intended dynamics may
be spoiled by the vast amount of accumulation of perturbations and
decoherence due to noise and undesired interactions.

To overcome this problem, several strategies have been proposed,
in which one promising solution to this problem is the shortcuts
to adiabaticity
dynamics~\cite{Demirplak2008,Berry2009,Chen2011,Chen2010,Dridi2009,Dridi2011,Campo2011,Campo2012,yehong16,yehongsr15,chenzhensr15,yehongpra151,yehongpra152,yehongpra153,lumeipra14,Pra08743402,Pra89053408}.
The basic idea of the shortcuts to adiabaticity dynamics is to
accelerate the dynamics towards the perfect final outcome so that
the accumulation of decoherence effects will be reduce
effectively. Based on this novel idea, several methods have been
proposed, such as counter-diabatic driving (equivalently, the
transitionless quantum
algorithm)~\cite{Demirplak2008,Berry2009,Campo2012,yehongsr15,lumeipra14},
Lewis-Riesenfeld inverse
engineering~\cite{Chen2011,yehongpra151,yehongpra152,yehongpra153,},
parallel adiabatic passage~\cite{Dridi2009}, and so on. Those
methods imply a strict time dependence of the pulses or detuning.
One can use the constructed composite pulses to improve
dramatically the fidelity of the adiabatic
passage~\cite{Demirplak2008,Berry2009,Chen2011,Chen2010,Dridi2009,Dridi2011,Campo2011,Campo2012,yehong16,yehongsr15,chenzhensr15,yehongpra151,yehongpra152,yehongpra153,lumeipra14,Pra08743402,Pra89053408}.
For example, based on the transitionless quantum driving, Chen
\textit{et al.} have proposed a shortcut to the STIRAP with
auxiliary pulses, which could provide a fast and robust approach
to population control~\cite{Chen2010}.  However, some dissipative
effects (e.g., the
dephasing)~\cite{Demirplak2002,Ivanov2004,Lacour2008,Lu2013,Issoufa2014,Graefe2013}
still produce significant negative effects on the schemes with
above methods. In addition, it is a pity that there has been a
long lack of studies on the impact on the imperfect initial
conditions which may occur in the experiment. Therefore, the
shortcuts to adiabaticity dynamics will also be limited severely
by the presence of the strong dissipation effects and imperfect
initial conditions.

Besides the shortcuts to adiabaticity dynamics, another popular
approach, so-called dissipative quantum dynamics
(DQD)~\cite{Verstraete2009,Kastoryano2011,Vacanti2009}, also work
well for coping with the dissipative effects. In fact, several
authors have pointed out that it is possible to generate
entanglement based on dissipative quantum dynamical
process~\cite{shao2012,shen2012,Wu2014}. The basic idea of the DQD
can be summarized as follows: the interaction of the system with
the environment is employed such that the target state becomes the
stationary state of the system. In other words, some specific
dissipative factors are no longer undesirable, but can be regarded
as the important resources. This idea is consistent with the
results in Ref.~\cite{Bender} where Bender has elaborated that a
$\mathcal{PT}$-symmetric Hamiltonian can produce a
faster-than-Hermitian evolution in a two-state quantum system,
while keeping the eigenenergy difference fixed. Recently, Torosov
\textit{et al.} have proposed a non-Hermitian generalization of
STIRAP, which allows one to increase speed and fidelity of the
adiabatic passage~\cite{Torosov2013}. However, the so-called
non-Hermitian shortcut is very sensitive to the initial
conditions, and the intermediate excited state always exists and
couldn't be neglected, which may be limited severely by the
presence of the dissipation effects of the excited state.  In
particular, Issoufa \textit{et al.} have found that the dephasing
effects will reduce the performance of the population transfer and
the fidelity can be far below the quantum computation target even
for small dephasing rates~\cite{Issoufa2014}. Thus, it is very
worthwhile to find an effective way to circumvent various
dissipative factors or the imperfect initial conditions which may
occur in the experiment.

In this paper, taking into account various decoherence factors
(such as the imperfect initial conditions, the dephasing and
damping effects, and the variations in the control parameters), we
aim at designing the pulses and detuning to increase the speed and
robustness of STIRAP by cutting off some coupling transitions via
the DQD. The scheme has the following advantages: (1) Unlike the
previous STIRAP or shortcut
schemes~\cite{firstSTIRAP,Bergmann1998,Vitanov2001,Demirplak2003,Giannelli2014,Sun2014,Demirplak2002,Ivanov2004,Lacour2008,Lu2013,Issoufa2014,Chen2010,Chen2011},
the dominant dissipative factors are no longer undesirable, in
fact, they are the important resources to the scheme. (2) By the
designed pulses and detuning, the target state will become the
stationary state of the system approximatively. Then, even though
the initial state is imperfect, the actual state will also follow
the evolution of the target state closely, thereby a complete
population transfer will be achieved fast and efficiently. (3) The
scheme is insensitive to the moderate fluctuations of experimental
parameters and the relatively large dissipation effects of the
excited state. (4) The approach makes it possible to improve the
speed of the system evolution in strong dissipative fields where
the standard STIRAP or shortcut schemes are invalid.

The rest of this paper is arranged as follows. In
Sec.~\ref{section:II}, we briefly analyse the dynamics of the
STIRAP process with the quantum jump
approach~\cite{Dalibard1992,Plenio1998,D2008}, the mechanism of
the population transfer in STIRAP and the limitation of
adiabaticity can be easily obtained. In Sec.~\ref{section:III},
taking into account the dephasing effects of the ground states,
the dynamics for the three-level system is also analysed with the
quantum jump approach. With the help of DQD, the pulses and
detuning design strategies are put forward. Then, we consider a
concrete three-level system example to show the usefulness of the
engineering method, the pulses and detuning shapes for the
constant dephasing rate are discussed in detail.
Sec.~\ref{section:IV} reports numerical analyses of the scheme.
Population engineering, the influences of fluctuations of
experimental parameters and the  dissipation effects of the
excited state, are discussed step by step. We conclude with a
summary of the scheme in Sec.~\ref{section:V}.

\section{The dynamics of the STIRAP process}\label{section:II}

\begin{figure}[htb]
\centering\scalebox{0.5}{\includegraphics{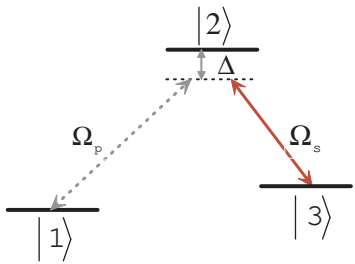}}
\caption{\label{fig1} Three-level $\Lambda$ system driven by two
coherent fields, a pump~(Stokes) field with Rabi frequency
$\Omega_{p}\ (\Omega_{s})$. The two fields have the same detuning
$\Delta$, where the time dependence has been omitted for
convenience.}
\end{figure}

In this section we briefly review the theory behind the STIRAP
process for the three-state systems. Let us consider a
single-photon detuning three-level quantum system in a $\Lambda$
configuration system as shown in Fig.~\ref{fig1}. The excited
state is labelled as $|2\rangle$ and two ground states are
$|1\rangle$ and $|3\rangle$. The transition between the levels
$|1\rangle\leftrightarrow|2\rangle$ is driven by the pump pulse
detuned from resonance by $\Delta(t)$ with the Rabi frequency
$\Omega_{p}(t)$. Whereas, the transition between the
$|2\rangle\leftrightarrow|3\rangle$ is driven by the Stokes pulse
detuned from resonance by $\Delta(t)$ with the Rabi frequency
$\Omega_{s}(t)$. The dipole transition
$|1\rangle\leftrightarrow|3\rangle$ is a forbidden transition. In
the absence of decoherence,  the dynamics of the system is
governed by the Schr\"{o}dinger equation $(\hbar=1)$
\begin{eqnarray}\label{eq0-1}
i\dot{c}(t)=H_{0}(t)c(t),
\end{eqnarray}
where the vector $c(t)$=$[c1(t),c2(t),c3(t)]^T$ contains the three
probability amplitudes of states $|1\rangle$, $|2\rangle$, and
$|3\rangle$. The Hamiltonian within the rotating wave
approximation~\cite{Scully1997} reads as
\begin{eqnarray}\label{eq0-2}
H_{0}(t)= \frac{1}{2}\left(\begin{array}{ccc}
 0            & \Omega_{p}(t) & 0\\
\Omega_{p}(t) & 2\Delta(t)    & \Omega_{s}(t)\\
 0            & \Omega_{s}(t) & 0\\
\end{array}\right).
\end{eqnarray}
The instantaneous  eigenvalues of the Hamiltonian $H_{0}(t)$ are
given by
\begin{eqnarray}\label{eq0-3}
\lambda_{+}(t)=\frac{\Omega_{0}(t)\cot{\phi}(t)}{2},~~
\lambda_{0}(t)=0,~~
\lambda_{-}(t)=\frac{-\Omega_{0}(t)\tan{\phi}(t)}{2},
\end{eqnarray}
where $\Omega_{0}(t)$ and the mixing angle $\phi(t)$ are,
respectively, defined by
\begin{eqnarray}\label{eq0-4}
\Omega_{0}(t)=\sqrt{\Omega_{p}(t)^2+\Omega_{s}(t)^2},~~
\tan\phi(t)=\frac{\Omega_{0}(t)}{\Delta(t)+\sqrt{\Delta(t)^2+\Omega_{0}(t)^2}},
\end{eqnarray}
with
\begin{eqnarray}\label{eq0-5}
\dot{\phi}(t)=\frac{\dot{\Omega}_{0}(t)\Delta(t)-\Omega_{0}(t)\dot{\Delta}(t)}{2(\Delta(t)^2+\Omega_{0}(t)^2)}.
\end{eqnarray}
Their corresponding eigenstates are given by the following
adiabatic states~\cite{Bergmann1998}
\begin{eqnarray}\label{eq0-6}
|a_{+}(t)\rangle&=&\sin{\theta}(t)\sin{\phi}(t)|1\rangle+\cos{\phi}(t)|2\rangle+\cos{\theta}(t)\sin{\phi}(t)|3\rangle,\cr
|a_{0}(t)\rangle&=&\cos{\theta}(t)|1\rangle-\sin{\theta}(t)|3\rangle,\cr
|a_{-}(t)\rangle&=&\sin{\theta}(t)\cos{\phi}(t)|1\rangle-\sin{\phi}(t)|2\rangle+\cos{\theta}(t)\cos{\phi}(t)|3\rangle,
\end{eqnarray}
where the mixing angle $\theta(t)$ is defined by
\begin{eqnarray}\label{eq0-7}
\tan\theta(t)=\frac{\Omega_{p}(t)}{\Omega_{s}(t)},\ \
\dot{\theta}(t)=\frac{\dot{\Omega}_{p}(t)\Omega_{s}(t)-\Omega_{p}(t)\dot{\Omega}_{s}(t)}{\Omega_{0}(t)^2}.
\end{eqnarray}

We can find that the eigenstate $|a_{0}(t)\rangle$ is a dark state
with zero projection on state $|2\rangle$, which is available for
the effective population transfer. It is advisable to project the
Schr\"{o}dinger equation into the so-called adiabatic states
$\{|a_{+}(t)\rangle,|a_{0}(t)\rangle,|a_{-}(t)\rangle\}$. The
probability amplitudes of the adiabatic states
$a(t)$=$[a_+(t),a_0(t),a_-(t)]^T$ are connected to the original
ones by using the transformation
\begin{eqnarray}\label{eq0-8}
c(t)={R}(t)a(t),
\end{eqnarray}
where the transformation matrix ${R}$ is given by
\begin{eqnarray}\label{eq0-9}
{R}(t)=\left(\begin{array}{ccc}
 \sin{\theta}(t)\sin{\phi}(t)   & \cos{\theta}(t)   & \sin{\theta}(t)\cos{\phi}(t)\\
\cos{\phi}(t)                   & 0                 & -\sin{\phi}(t)\\
 \cos{\theta}(t)\sin{\phi}(t)   &-\sin{\theta}(t)   & \cos{\theta}(t)\cos{\phi}(t)\\
\end{array}\right).
\end{eqnarray}
The Schr\"{o}dinger equation in the adiabatic basis can be written
as
\begin{eqnarray}\label{eq0-10}
i\dot{a}(t)=H^{a}_0a(t),
\end{eqnarray}
with $H^{a}_0=R(t)^{-1}H_{0}(t)R(t)-iR(t)^{-1}\dot{R(t)}$  or,
explicitly,
\begin{eqnarray}\label{eq0-11}
H^{a}_0=\left(\begin{array}{ccc}
\lambda_{+}(t)     &\Omega^{*}_1(t)  & \Omega^{*}_3(t)\\
\Omega_1(t)        & 0               & \Omega^{*}_2(t)\\
\Omega_3(t)        & \Omega_2(t)      & \lambda_{-}(t)\\
\end{array}\right),
\end{eqnarray}
where
\begin{eqnarray}\label{eq0-12}
\Omega_1(t)=-i\dot{\theta}(t)\sin\phi(t),
\Omega_2(t)=i\dot{\theta}(t)\cos\phi(t),
\Omega_3(t)=-i\dot{\phi}(t).
\end{eqnarray}

\begin{figure}[htb]
\centering\scalebox{0.5}{\includegraphics{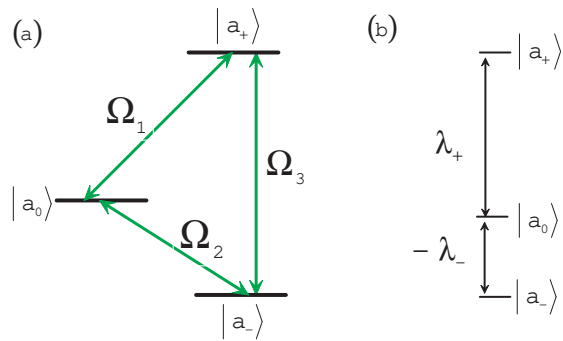}}
\caption{\label{fig2} (a) Effective level scheme of the STIRAP
process driven by three effective coherent fields $\Omega_1(t)$,
$\Omega_2(t)$, and $\Omega_3(t)$. (b) The energy difference
between the adiabatic states $|a_+(t)\rangle$ and $|a_0(t)\rangle$
 ($|a_0(t)\rangle$ and $|a_-(t)\rangle$) is given by $\lambda_{+}$
($-\lambda_{-}$).}
\end{figure}

In order to gain a qualitative understanding of the STIRAP
process, we now use the quantum jump
approach~\cite{Dalibard1992,Plenio1998,D2008} to analyse current
system as shown in Fig.~\ref{fig2}. We can easily find that the
STIRAP process can be considered as a three-level toy model driven
by three effective coherent fields $\Omega_1(t)$, $\Omega_2(t)$,
and $\Omega_3(t)$. When
\begin{eqnarray}\label{eq0-13}
|\Omega_1(t)|\ll|\lambda_{+}(t)|,
|\Omega_2(t)|\ll|\lambda_{-}(t)|,
|\Omega_3(t)|\ll|\lambda_{+}(t)-\lambda_{-}(t)|,
\end{eqnarray}
the transitions between the adiabatic states will be greatly
suppressed due to the rapid oscillations. In other words, when the
Eq.~(\ref{eq0-13}) is satisfied, if the system is initially in one
of the adiabatic states, then it stays in the adiabatic state
during the evolution. More specifically, the initially populated
state is $|\psi(0)\rangle=|1\rangle$, and it coincides with
$|a_0(t)\rangle$ for ${\theta}(t)\approx$0, which can be realized
with the counterintuitive pulses
order~\cite{Bergmann1998,Vitanov2001,Demirplak2003,Giannelli2014,Laine1996}.
For a counterintuitive order of the pump and Stokes pulses, we
have the relations
\begin{eqnarray}\label{eq0-14}
\tan\theta(t)\stackrel{t\rightarrow t_i}{\longrightarrow}0,
\tan\theta(t)\stackrel{t\rightarrow t_f}{\longrightarrow}\infty,
\end{eqnarray}
where $t_i$ ($t_f$) is the initial (final) time for the
counterintuitive pulses. Then, as time evolves, the mixing angle
$\theta(t)$ rises from 0 to $\pi/2$, and we achieve a complete
population transfer
\begin{eqnarray}\label{eq0-15}
|a_0(t_i)\rangle=|1\rangle\longrightarrow|a_0(t_f)\rangle=-|3\rangle.
\end{eqnarray}
Here, we should note that above derivations are based on the ideal
cases, namely perfect adiabaticity, perfect initial state, and
without the decoherence effects. However, the perfect adiabaticity
in Eq.~(\ref{eq0-13}) always means long evolution time which is
never experimentally possible nor theoretically expected.
Therefore, some nonadiabatic couplings $\Omega_1(t)$ and
$\Omega_2(t)$ are always present, which limits the efficiency of
STIRAP. Furthermore, the imperfect of initial state and the
decoherence effects also have negative effects on the system.
Another point which is important to stress is that the coupling
term $\Omega_3(t)$ between $|a_+(t)\rangle$ and $|a_-(t)\rangle$,
seems not necessary for a full passage from $|1\rangle$ to
$|3\rangle$ in principle. For the purpose of convenience, in
several of the published numerical calculations of the STIRAP
process~\cite{Sun2014,Torosov2013,Giannelli2014}, $\dot{\phi}(t)$
are vanished in the interesting case of $\Delta(t)=\eta
\Omega_{0}(t)$. Here, $\eta$ can be any constant, also zero,
leading to the convenient choice $\Delta(t)=0$. However, in the
scheme, the detuning $\Delta(t)$  plays a important role in the
presence of dephasing effects and imperfect initial conditions.

\section{Dissipative quantum dynamics shortcut}\label{section:III}

According to the DQD, the decoherence effects in the system should
be employed as far as possible to improve the STIRAP process. For
the standard STIRAP process, the dissipation effects of the ground
states may be the dominant source of decoherence, since the system
always stays in $|a_0(t)\rangle$ which is constituted by two
ground states  during the evolution. Therefore, in the following
discussion, we will concentrate on the dissipation effects of the
ground states. Other dissipation effects including the radiation
processes and the dephasing effect of the excited state
$|2\rangle$, are considered as the perturbations, we will discuss
their impacts on the scheme in the Sec.~\ref{section:IV}.

In general, for the dissipative quantum systems, the complex
energies with negative imaginary parts are used to describe an
overall probability decrease that models decay, transport, or
scattering phenomena (see, e.g.,
Refs.~\cite{Dattoli1990,Moiseyev1998,Okolowicz2003,Berry2004} and
references therein). Here, without loss of generality, we consider
the dissipation effects of the ground states with the
form~\cite{Torosov2013,Randall2015}
\begin{eqnarray}\label{eq1-0}
H_{I}=-i\frac{\Gamma(t)}{2}[|1\rangle\langle1|-|3\rangle\langle3|],
\end{eqnarray}
where $-i\Gamma(t)/2$ correspond to time-varying complex energies
of the bare states $|1\rangle$ and $|3\rangle$, which may be
induced by environmental effects or the fluctuating frequencies of
the control fields. In fact, similar dissipation effects are
widely discussed in
Refs.~\cite{Demirplak2002,Ivanov2004,Lacour2008,Lu2013,Issoufa2014,Sun2014,Giannelli2014},
and such dissipation effects can be considered as the dephasing
effects which occur in the experiment due to the collisions or
phase fluctuations of the control fields. In the following, we
will show how to achieve a complete population transfer in a short
time with such ``dephasing effects''.

\subsection{Pulses and detuning design strategy}

Taking into account the dephasing effects of the ground states,
the Hamiltonian of the system can be rewritten as
\begin{eqnarray}\label{eq1-1}
H_{\Gamma(t)}=H_{0}+H_{I}=\frac{\hbar}{2}\left(\begin{array}{ccc}
 -i\Gamma(t)  & \Omega_{p}(t) & 0            \\
\Omega_{p}(t) & 2\Delta(t)    & \Omega_{s}(t)\\
 0            & \Omega_{s}(t) & i\Gamma(t)   \\
\end{array}\right).
\end{eqnarray}
In the adiabatic basis Eq.~(\ref{eq0-6}), this Hamiltonian has the
form
\begin{eqnarray}\label{eq1-2}
H^{a}_{\Gamma(t)}= \left(\begin{array}{ccc}
\Omega_{++}(t)    & \Omega_{+0}(t)  & \Omega_{+-}(t)\\
\Omega_{0+}(t)    & \Omega_{00}(t)  & \Omega_{0-}(t)\\
\Omega_{-+}(t)    & \Omega_{-0}(t)  & \Omega_{--}(t)\\
\end{array}\right),
\end{eqnarray}
where
\begin{eqnarray}\label{eq1-3}
\Omega_{++}(t)&=&\lambda_{+}(t)+i\frac{\Gamma\cos2\theta(t)\sin^2{\phi(t)}}{2},
~\Omega_{+0}(t)=i\dot{\theta}(t)\sin\phi(t)-i\frac{\Gamma\sin2\theta(t)\sin{\phi(t)}}{2},\cr
\Omega_{+-}(t)&=&i\dot{\phi}(t)+i\frac{\Gamma\cos2\theta(t)\sin2{\phi(t)}}{4},
~\Omega_{0+}(t)=-i\dot{\theta}(t)\sin\phi(t)-i\frac{\Gamma\sin2\theta(t)\sin{\phi(t)}}{2},\cr
\Omega_{00}(t)&=&-i\frac{\Gamma\cos2\theta(t)}{2},
~~~~~~~~~~~~~~~~~~~~~~~~\Omega_{0-}(t)=-i\dot{\theta}(t)\cos\phi(t)-i\frac{\Gamma\sin2\theta(t)\cos{\phi(t)}}{2},\cr
\Omega_{--}(t)&=&\lambda_{-}(t)+i\frac{\Gamma\cos2\theta(t)\cos^2{\phi(t)}}{2},
\Omega_{-0}(t)=i\dot{\theta}(t)\cos\phi(t)-i\frac{\Gamma\sin2\theta(t)\cos{\phi(t)}}{2},\cr
\Omega_{-+}(t)&=&-i\dot{\phi}(t)+i\frac{\Gamma\cos2\theta(t)\sin2{\phi(t)}}{4}.
\end{eqnarray}

\begin{figure}[htb]
\centering\scalebox{0.5}{\includegraphics{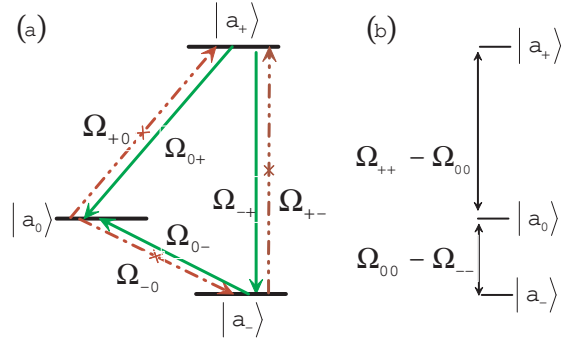}}
\caption{\label{fig3} (a) The resultant energy level diagram of
the three-level $\Lambda$ system when the dissipation effects of
the ground states [see Eq.~(\ref{eq1-0})] are considered. (b) The
energy difference between the adiabatic states $|a_+(t)\rangle$
and $|a_0(t)\rangle$ ($|a_0(t)\rangle$ and $|a_-(t)\rangle$) in
this case is given by $\Omega_{++}-\Omega_{00}$
 ($\Omega_{00}-\Omega_{--}$).}
\end{figure}
Similar to the analysis procedure in Sec.~\ref{section:II}, we
will also use the quantum jump approach to analyse the current
model. The resultant energy level diagram is shown in
Fig.~\ref{fig3}. Interestingly, we find that the effective
couplings  between the adiabatic states are different from each
other. Enlightened by the idea of DQD, we aim at designing the
pulses and detuning to cut off some undesirable coupling
transitions so that $|a_0(t)\rangle$ becomes the stationary state
of the system approximately. Then, even though the initial state
is imperfect, the actual state also will follow the evolution of
$|a_0(t)\rangle$ closely, thereby transferring the population from
$|1\rangle$ to $|3\rangle$ will be achieved efficiently in a short
time. According to the DQD, following couplings
\begin{eqnarray}\label{eq1-4}
\Omega_{+0}(t), \Omega_{-0}(t), \Omega_{-+}(t), \Omega_{+-}(t),
\end{eqnarray}
deviate from our goal, thus, we should make them vanish as much as
possible. We can find that $\Omega_{\pm0}(t)$ can be vanished by
setting
\begin{eqnarray}\label{eq1-5}
\frac{2\dot{\theta}(t)}{\sin2\theta(t)}=\Gamma(t)
=\frac{\dot{\Omega}_{p}(t)}{\Omega_{p}(t)}-\frac{\dot{\Omega}_{s}(t)}{\Omega_{s}(t)}
=\frac{d}{dt}\ln\frac{{\Omega}_{p}(t)}{\Omega_{s}(t)},
\end{eqnarray}
however, $\Omega_{-+}(t)$ and $\Omega_{+-}(t)$ could not be
vanished simultaneously. The next best thing is to nullify one of
them so that decreasing the coupling oscillation between the
undesirable states $|a_+(t)\rangle$ and $|a_-(t)\rangle$, thereby
shortening the evolution time as much as possible. Without loss of
generality,  we can set $\Omega_{+-}(t)$=0, then $\Delta(t)$
should satisfy following equation
\begin{eqnarray}\label{eq1-6}
{\dot{\Delta}(t)}=\frac{\dot{\Omega}_{0}(t)}{\Omega_{0}(t)}{\Delta}(t)+\Gamma(t){\cos2\theta(t)}.
\end{eqnarray}
Therefore, the detuning $\Delta(t)$  is designed to cut off the
undesirable coupling transitions which plays an important role in
the scheme.

Up to now, we have in principle designed the pulses
${\Omega_{p,s}(t)}$ and the detuning $\Delta(t)$ based on the DQD.
Here, we should make some remarks on the design strategies. (1)
Generally, the dephasing rate $\Gamma(t)$ is uncontrollable which
naturally occurs as a result of collisions or fluctuations of the
fields. However, in some cases, $\Gamma(t)$ can be controlled as
an effective decay rate by further interactions, see, e.g.,
Ref.~\cite{Muga2008}. Therefore,  if the form of $\Gamma(t)$ is
fixed, the pulses ${\Omega_{p,s}(t)}$ can be derived  with reverse
thinking according to the Eq.~(\ref{eq1-5}). (2) Once the pulses
are fixed, which means the detuning $\Delta(t)$ is fixed according
to the Eq.~(\ref{eq1-6}). However, we may not be able to get its
analytical solution directly from Eq.~(\ref{eq1-6}), if the forms
of pulses and the dephasing rate are complicated. In this case,
the fitting of numerical solution should be applied. Notice that
Eq.~(\ref{eq1-5}) and Eq.~(\ref{eq1-6}) are the primary results to
be used in following work.

\subsection{The pulses and the detuning shapes}

The scheme is based on the DQD, the dissipative factors play a key
role. By ``dissipative factors" we mean here a slight deviation of
the population with respect to the initial state in the ideal
protocol and the dephasing effects of the ground states. In the
following, we will take a concrete example to show the usefulness
of the dissipative quantum dynamics shortcut design strategy with
different dissipative factors. For convenient discussion, we
consider a realistic case of a small population in the bare states
$|2\rangle$ and $|3\rangle$ for the initial state
\begin{eqnarray}\label{eq2-1}
|\psi(0)\rangle=\sqrt{1-2\epsilon^2}|1\rangle+\epsilon(-|2\rangle+|3\rangle),
\end{eqnarray}
where $\varepsilon$ is a small deviation constant. Furthermore, we
shall take $\Gamma(t)$ as a constant $\Gamma$ in the Markovian
description which was widely discussed in
Refs.~\cite{Sun2014,Giannelli2014,Dattoli1990,Moiseyev1998,Okolowicz2003,Berry2004},
although, in a general non-Markovian
case~\cite{Yupra1999,Jing2010,Maniscalco2006}, it could also
depend on time. For the constant dephasing rate $\Gamma$, it can
easily be shown that the pulses satisfy following equation, taking
into account Eq.~(\ref{eq1-5}),
\begin{eqnarray}\label{eq2-2}
\frac{{\Omega}_{p}(t)}{\Omega_{s}(t)}=C\exp({\Gamma t}),
\end{eqnarray}
where $C$ is a arbitrary non-zero constant. In addition, according
to the standard STIRAP technique, the counterintuitive order of
the pump and Stokes pulses can be written
as~\cite{Bergmann1998,Vitanov2001,Demirplak2003,Giannelli2014,Laine1996}
\begin{eqnarray}\label{eq2-3}
\Omega_{p}(t)=\Omega_{\textrm{peak}}f(\frac{t-\tau_{0}/2}{T}),
\Omega_{s}(t)=\alpha\Omega_{\textrm{peak}}f(\frac{t+\tau_{0}/2}{T}),
\end{eqnarray}
where $\Omega_{\textrm{peak}}$ is the peak Rabi frequency, $T$ is
the pulse width,  and $\alpha$ is a non-zero scaling parameter,
while $\tau_{0}$ is the delay between pulses which is imposed
$\tau_{0}>$0 by the counterintuitive sequence condition. We can
find the simplest solutions of Eq.~(\ref{eq2-2}) and
Eq.~(\ref{eq2-3}) for the pulses ${\Omega_{p,s}(t)}$ are
\begin{eqnarray}\label{eq2-4}
\Omega_{p}(t)=\Omega_{\textrm{peak}}\exp({-[({t-\tau_{0}/2})/{T}]^2}),
\Omega_{s}(t)=\Omega_{\textrm{peak}}\exp({-[({t+\tau_{0}/2})/{T}]^2}),
\end{eqnarray}
where $\tau_{0}=\Gamma T^2/2$, $C$ and $\alpha$ are assumed as 1
for simplicity.  In the following we will scale all parameters
with respect to the width of the pulses,
\begin{eqnarray}\label{eq2-5}
\Omega_{\textrm{peak}}=\frac{\Omega}{{T}},
\Gamma=\frac{\gamma}{{T}},  \tau_{0}=\frac{\gamma T}{2}=\tau{T},
\end{eqnarray}
so that the system evolution is characterized by a time-scale
invariance in the numerical calculations.

\begin{figure}[htb]
\centering\scalebox{0.35}{\includegraphics{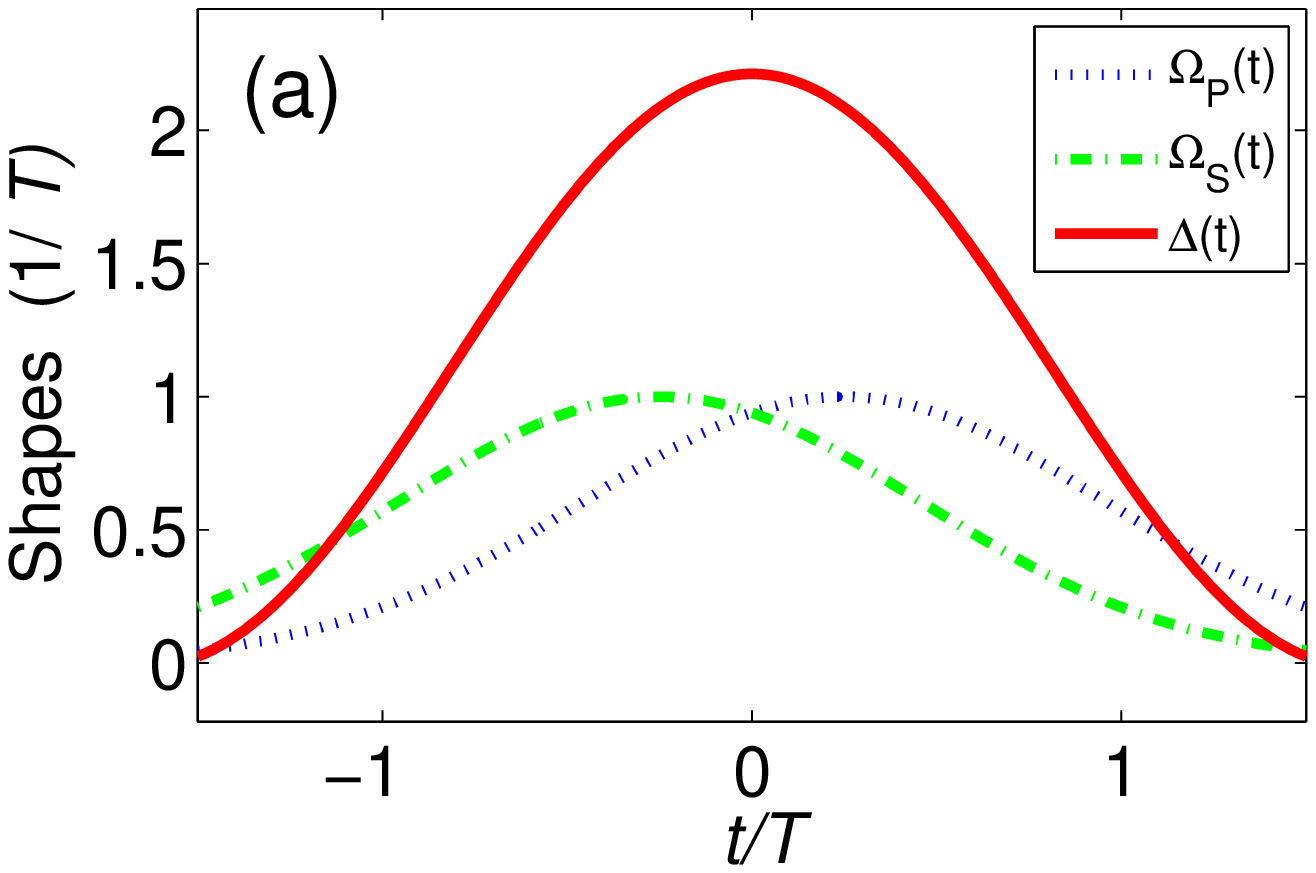}\includegraphics{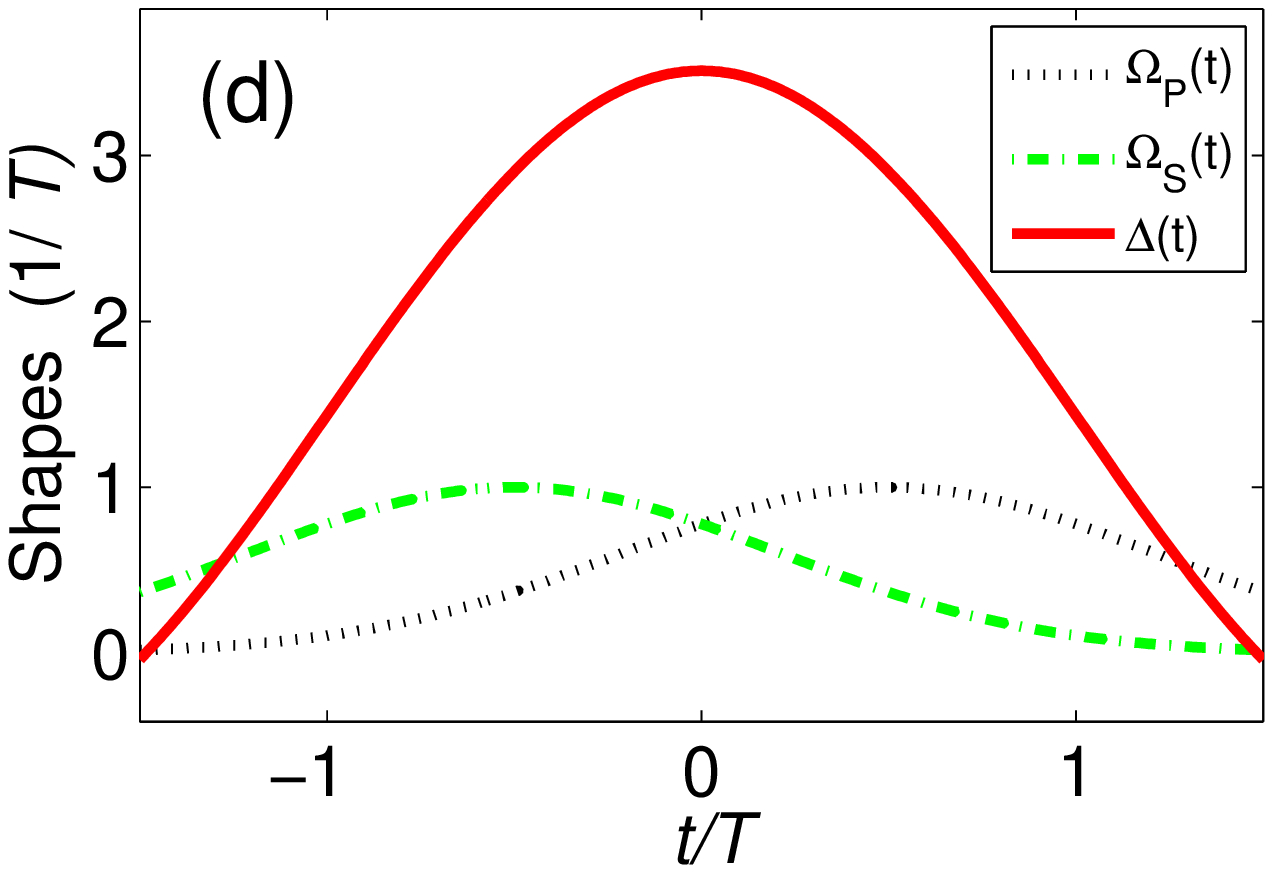}}
\scalebox{0.35}{\includegraphics{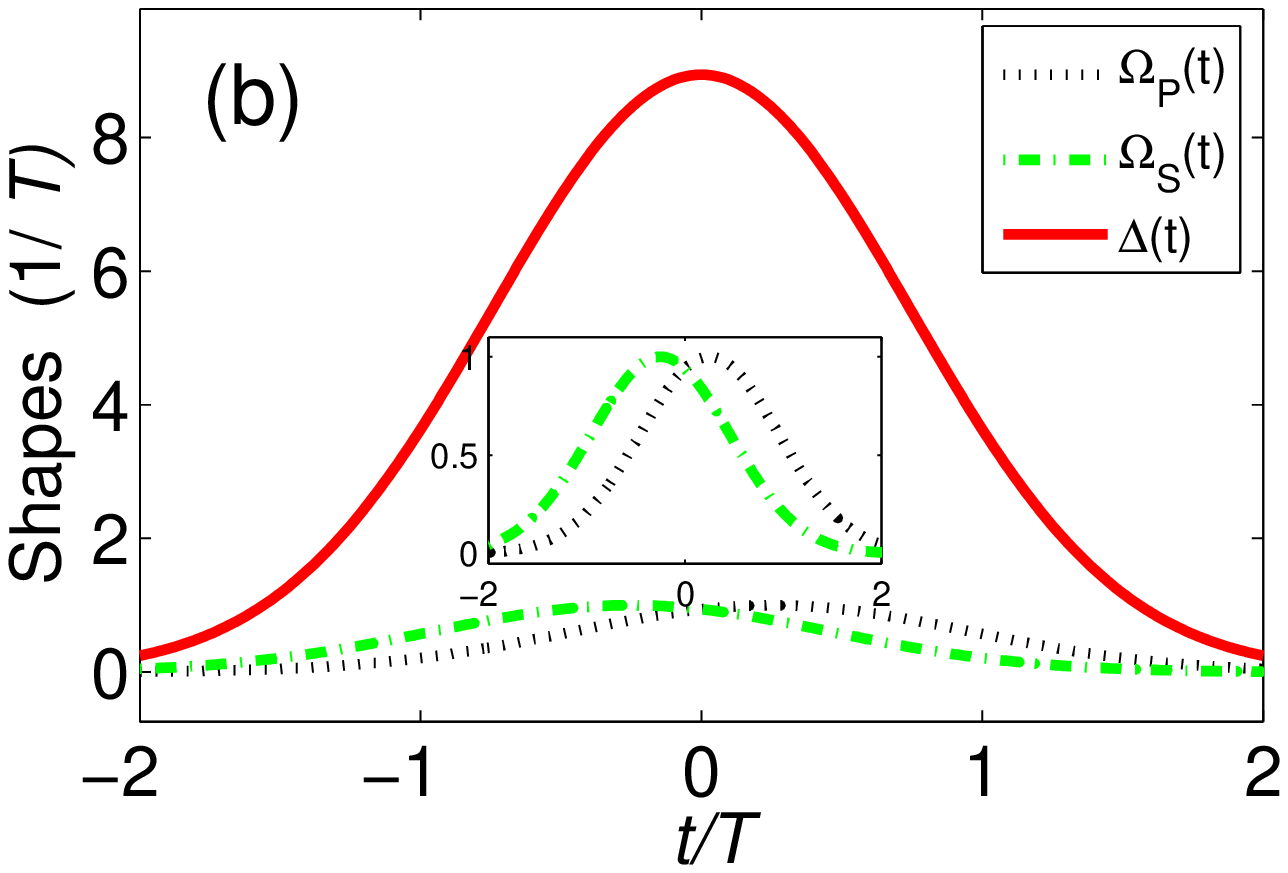}\includegraphics{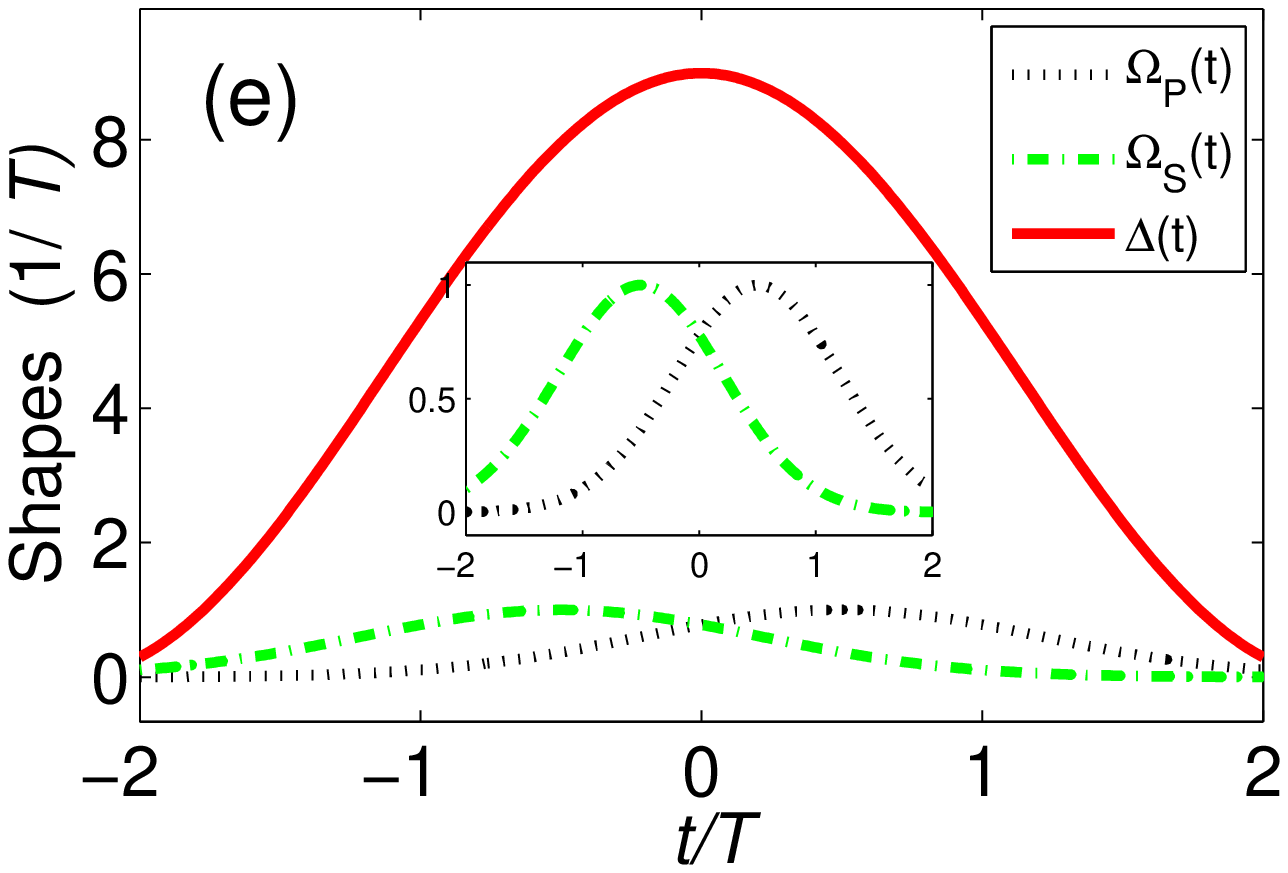}}
\scalebox{0.35}{\includegraphics{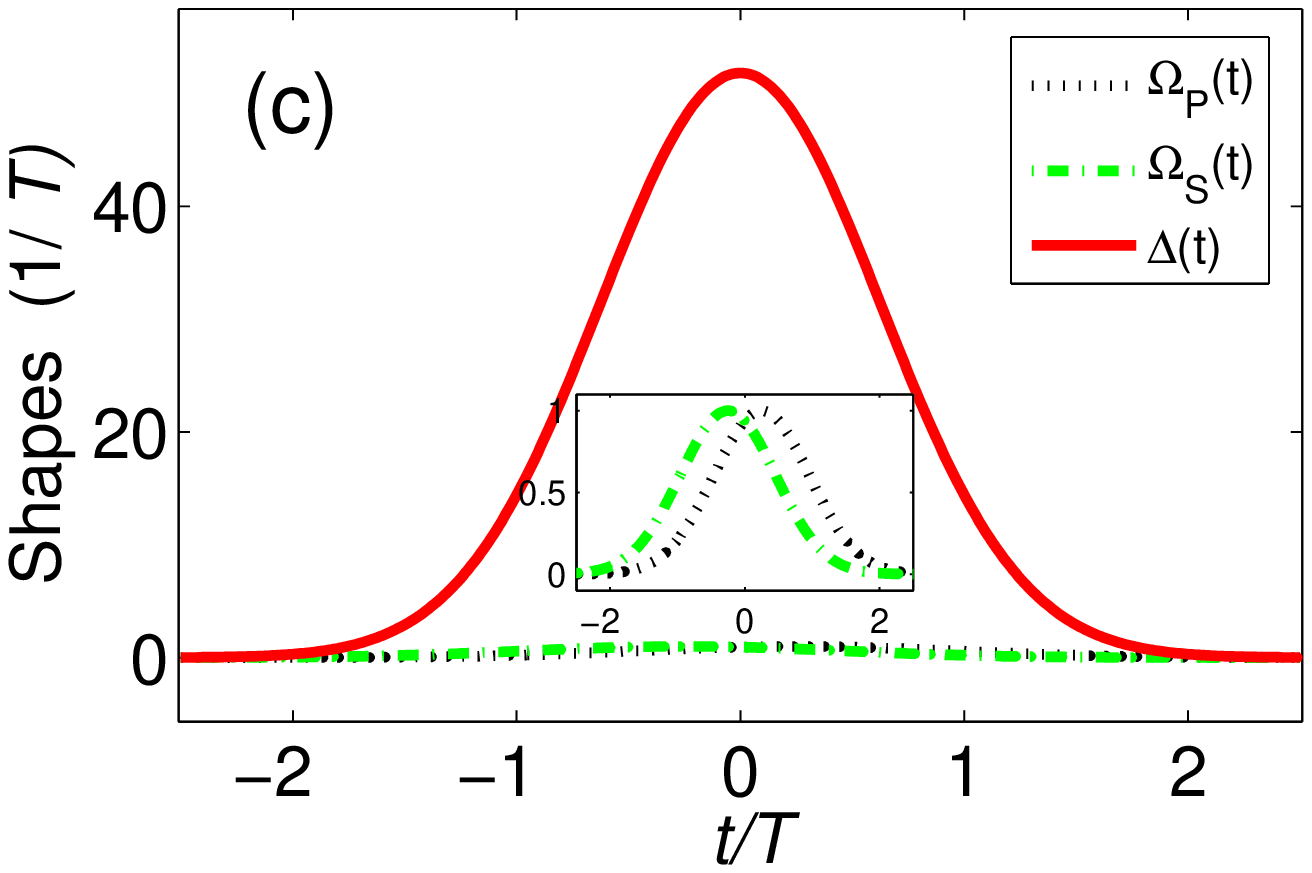}\includegraphics{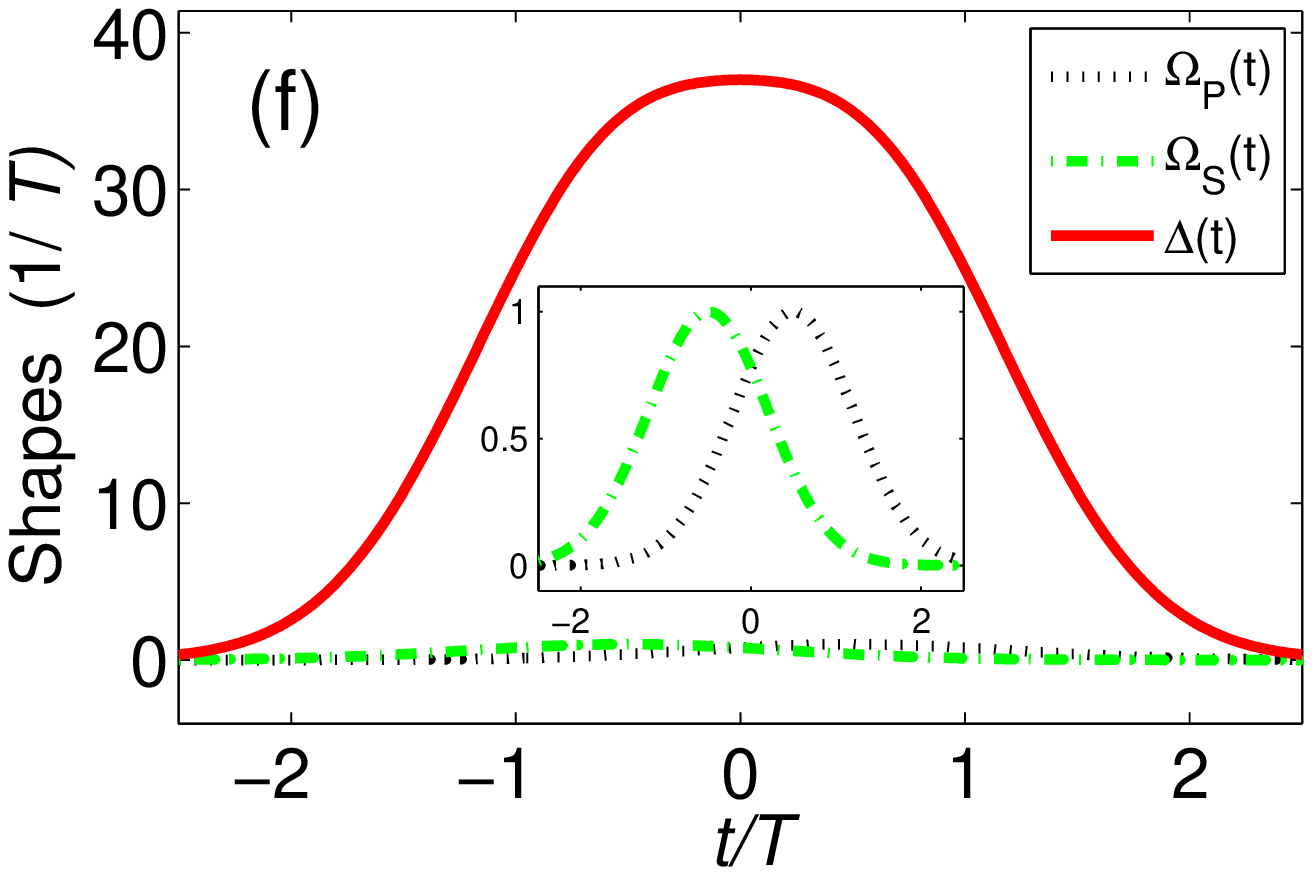}}
\caption{\label{fig4} Evolution in time of the pump and Stokes
pulses and the detuning  with different dephasing rates within
different STIRAP time intervals. For (a), (b) and (c) the
dephasing rate $\gamma=1$: (a) $t_f=1.5T$, (b) $t_f=2T$, (c)
$t_f=2.5T$. For (d), (e) and (f) the dephasing rate $\gamma=2$:
(d) $t_f=1.5T$, (e) $t_f=2T$, (f) $t_f=2.5T$. The shapes of the
pump and Stokes pulses are based on Eq.~(\ref{eq2-4}) with
$\Omega=1$, while the detuning $\Delta(t)$ are ploted according to
Eq.~(\ref{eq1-6}) by numerical calculations with the boundary
condition $\Delta(t_i)=\Delta(t_f)=0$. }
\end{figure}

\begin{table}
\newcommand{\tabincell}[2]{\begin{tabular}{@{}#1@{}}#2\end{tabular}}
\centering \caption{\label{table1}Formal solution for the detuning
$\Delta_n(t)~(n=a,b,c,d,e,f)$, where $\Delta_n(t)$ is the detuning
in the Fig.~\ref{fig4}(n). The units for $\Delta_{0,1}$ and
$\omega$ are $1/T$, while that for $\mu_{0,1}$ and $\nu_{0,1}$ are
$T$.}
\begin{tabular}{c|c|c|c|c|c|c|c}
\hline\hline \multicolumn{1}{|c|}{\tabincell{c}{Fitting function}}
&\multicolumn{1}{|c|}{Detuning}
&\multicolumn{1}{|c|}{Parameters}  \\
\hline
\multicolumn{1}{|c|}{\tabincell{c}{\tabincell{c}{Fourier\\$\Delta_{0}+\Delta_{1}\cos\omega{t}$}}}
&\multicolumn{1}{|c|}{\tabincell{c}{$\Delta_{a}(t)$\\$\Delta_{d}(t)$\\$\Delta_{e}(t)$}}
&\multicolumn{1}{|c|}{\tabincell{c}{$\Delta_{0}=\Delta_{1}=1.12, \omega=1.92$ \\
$\Delta_{0}=1.43,\Delta_{1}=2.09, \omega=1.57$ \\ $\Delta_{0}=\Delta_{1}=4.49, \omega=1.39$}}\\
\hline
\multicolumn{1}{|c|}{\tabincell{c}{\tabincell{c}{Gaussian\\$\Delta_{0}e^{-[({t-\mu_{0}})/{\nu_{0}}]^2}$\\
+$\Delta_{1}e^{-[({t-\mu_{1}})/{\nu_{1}}]^2}$}}}
&\multicolumn{1}{|c|}{\tabincell{c}{$\Delta_{b}(t)$\\$\Delta_{c}(t)$\\$\Delta_{f}(t)$}}
&\multicolumn{1}{|c|}{\tabincell{c}{$\Delta_{0}=8.94, \nu_{0}=1.92, \Delta_{1}=\mu_{0,1}=\nu_{1}=0$\\
$\Delta_{0}=51.83, \nu_{0}=0.88, \Delta_{1}=\mu_{0,1}=\nu_{1}=0$\\
$\Delta_{0}=\Delta_{1}=28.85, \mu_{0}=-\mu_{1}=0.6, \nu_{0,1}=0.9$}}\\
\hline
\end{tabular}
\end{table}

Now we start to study the dynamics of the detuning $\Delta(t)$
with different dephasing rates. In Fig.~\ref{fig4}, we plot the
evolution in time of the pump and Stokes pulses and the detuning
with different dephasing rates within different STIRAP time
intervals, where we have assumed that the STIRAP time interval is
symmetric, $t_f$=$-t_i$. In the examples of Fig.~\ref{fig4}, we
can find the shapes of the pump and Stokes pulses and the detuning
$\Delta(t)$ are very simple which are feasible for the current
experimental
techniques~\cite{Bason2012,Zhang2013,Du2016,Campo2016}. From an
experimental view point, we give the formal solution for the
detuning $\Delta(t)$ with some simple functions by curve fitting
as shown in Table~\ref{table1}. Moreover, we can find the peak of
the detuning $\Delta(t)$ will signally increase with the increase
of the STIRAP time interval, and the maximum value of the peak of
$\Delta(t)$ is about $51/T$ in the Fig.~\ref{fig4}(c). However,
the shapes of the detuning $\Delta(t)$ only change slightly for
different dephasing rates within the same STIRAP time interval. Up
till now, we have successfully designed the pulses and the
detuning for different dephasing rates within different STIRAP
time intervals.

\section{ Numerical analyses}\label{section:IV}

\begin{figure}[htb]
\centering\scalebox{0.35}{\includegraphics{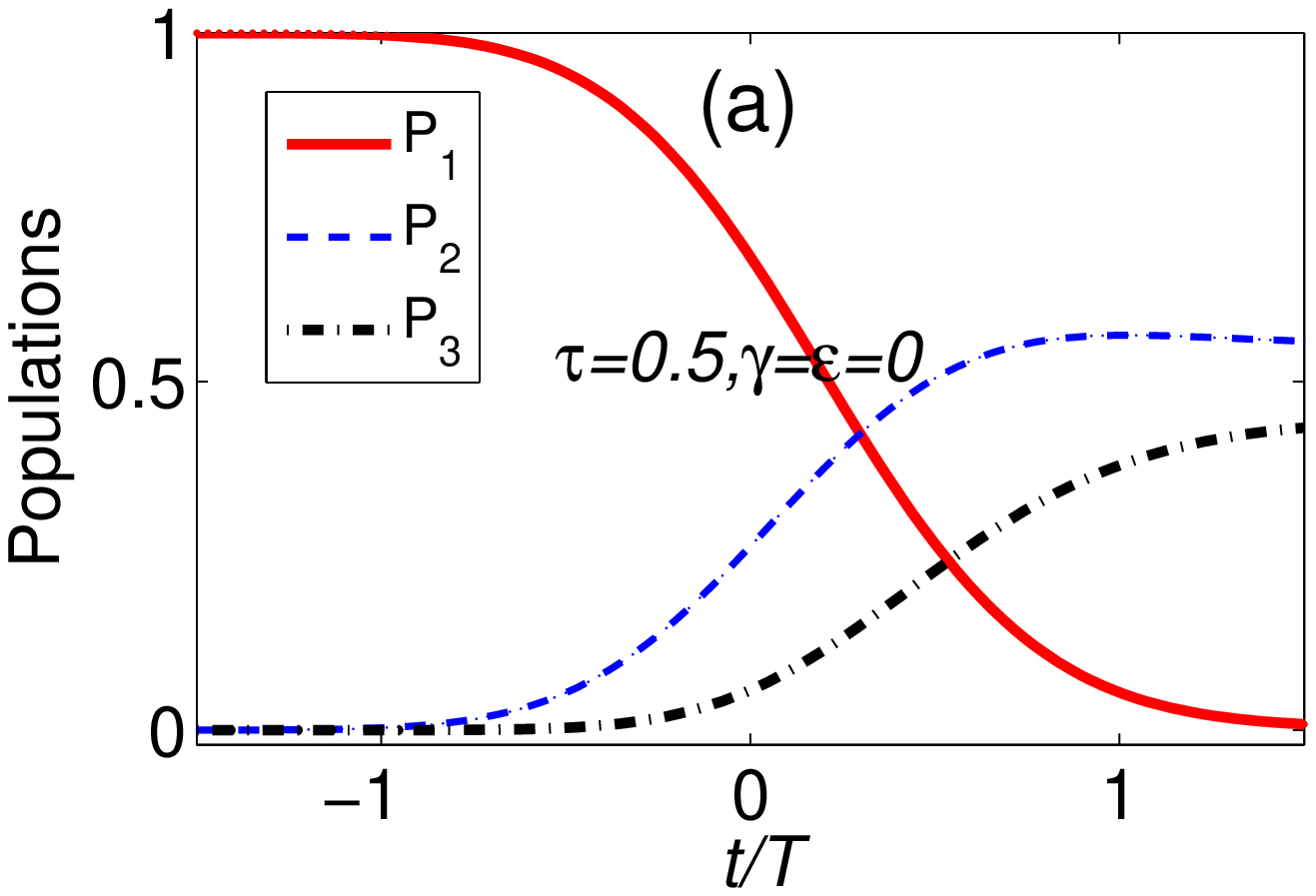}\includegraphics{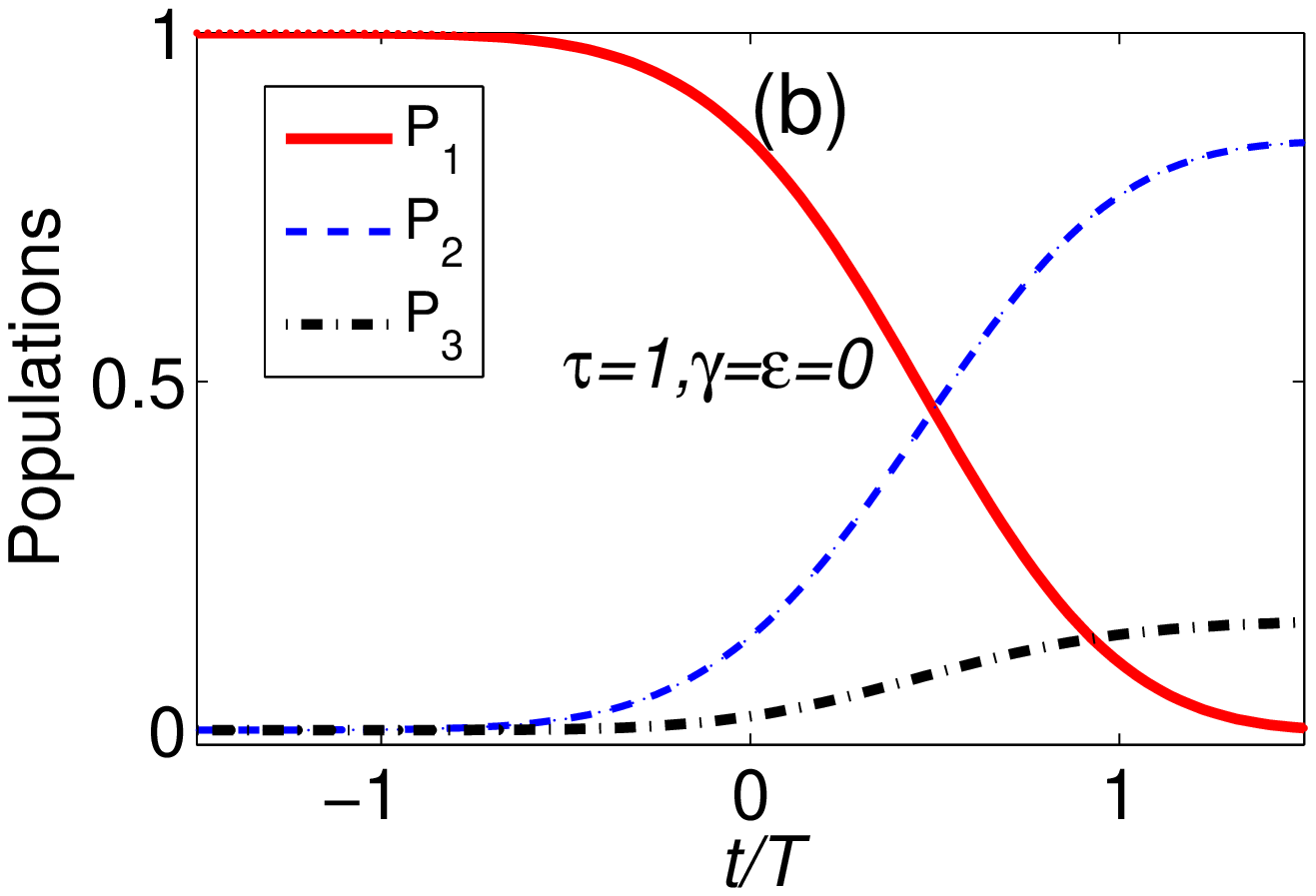}}
\centering\scalebox{0.35}{\includegraphics{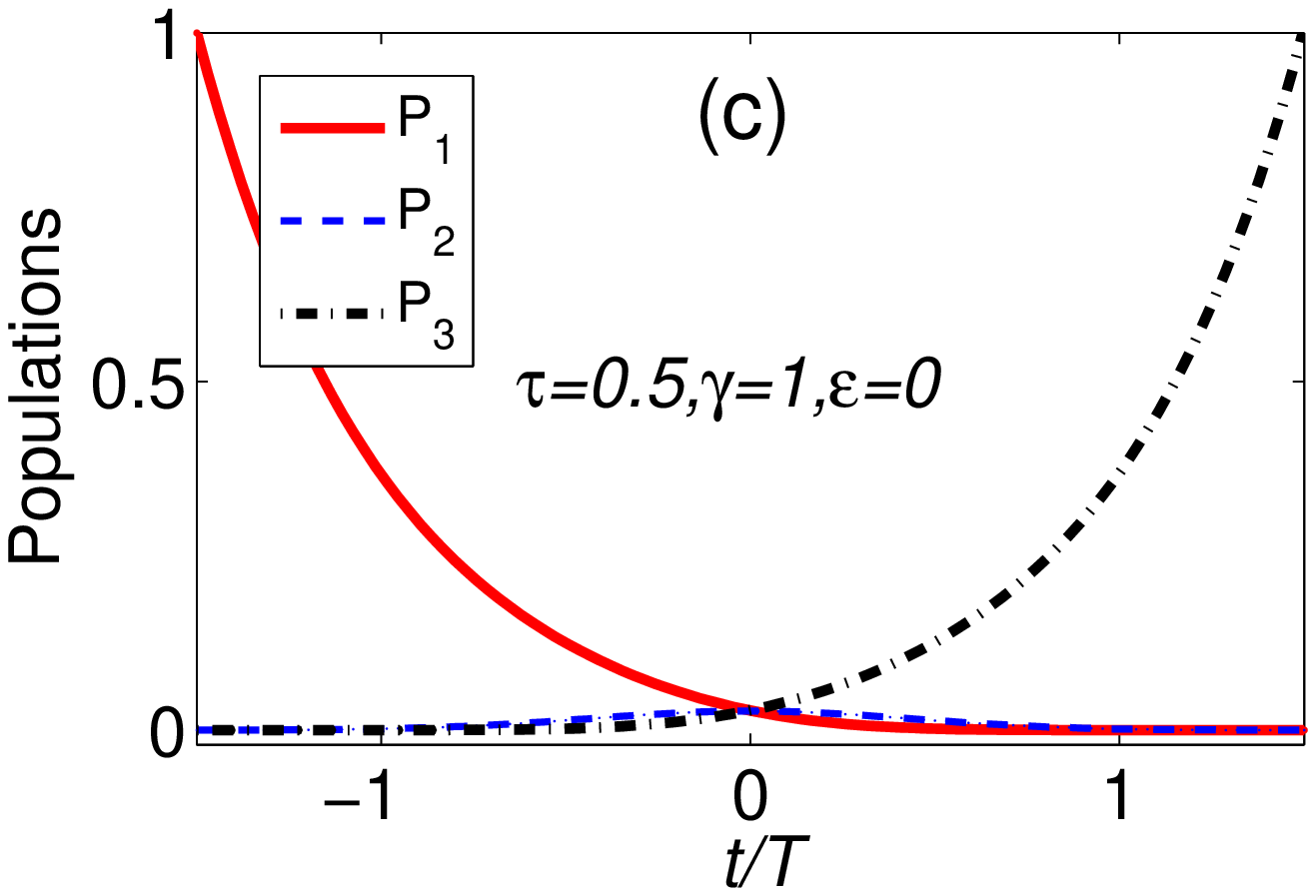}\includegraphics{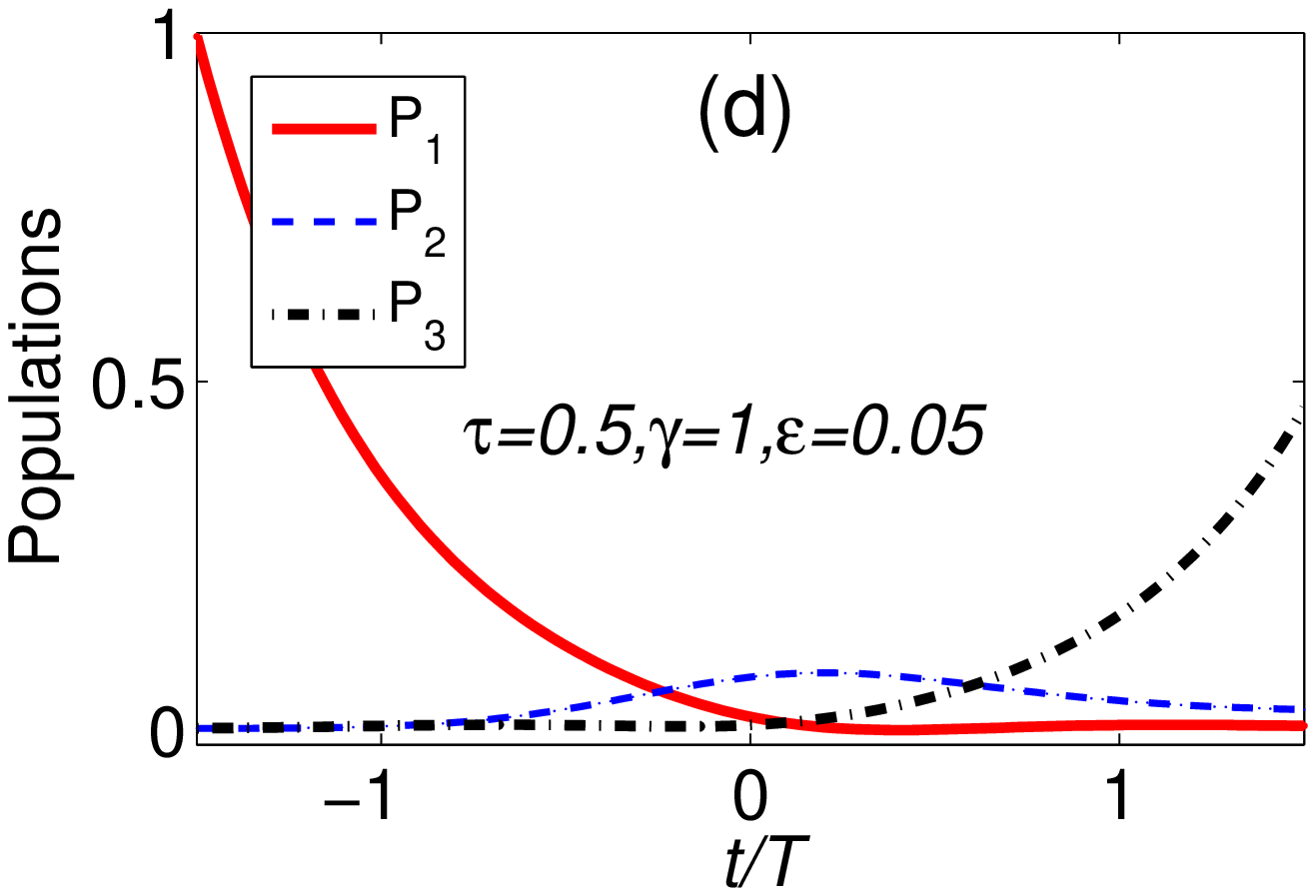}}
\caption{\label{fig5}  The time evolution of population $P_1$
($P_2$, $P_3$) for the bare state $|1\rangle$ ($|2\rangle$,
$|3\rangle$) with the pump and Stokes pulses  with different
imperfect initial states for the traditional STIRAP process without
detuning. The pump and Stokes pulses are based on Eqs. (26-27),
while the initial state is based on Eq. (23). For (a) and (b),  the
initial states are perfectly populated in $|1\rangle$ and the
dephasing effects are ignored, the delays $\tau_0$ between pulses
are 0.5 and 1, respectively.  For (c) and (d), the dephasing rates
$\gamma$ are 1 and the delays $\tau_0$ between pulses are $0.5$, the
initial state in (c) is perfectly populated in $|1\rangle$, whereas
there exists a slight deviation $\varepsilon=0.05$ from $|1\rangle$
in (d).}
\end{figure}

Before the elaborating on the performance of the scheme for the
complete population transfer from $|1\rangle$ to $|3\rangle$, we
will briefly evaluate the performance of traditional STIRAP
process with the pump and Stokes pulses which have been shown in
Fig.~\ref{fig4} with the help of population engineering. The
population for the bare state $|i\rangle~(i=1,2,3)$ is given
through the relation $P_{i}=|\langle{i}|\rho(t)|i\rangle|$, where
$\rho(t)$ is the density operator of the system at the time $t$.
For convenient discussion, we plot the time evolution of
populations for the bare states $|1\rangle$, $|2\rangle$,
$|3\rangle$ with four sets of parameters
$\{\tau=0.5,\gamma=\epsilon=0\}$, $\{\tau=1,\gamma=\epsilon=0\}$,
$\{\gamma=1,\tau=0.5,\epsilon=0\}$,
$\{\gamma=1,\tau=0.5,\epsilon=0.05\}$ in Fig.~\ref{fig5}. As shown
in Figs.~\ref{fig5}(a)-(b), we can find the complete population
transfer fails, even though the initial state is perfectly
populated in $|1\rangle$ and the dephasing effects are ignored.
Figure~\ref{fig5}(c) shows a complete population transfer when the
initial state is perfectly populated on $|1\rangle$ and there
exists the dephasing effect $\gamma=1$. However, the intermediate
state $|2\rangle$ always exists and couldn't be neglected, which
may be severe if the excited state $|2\rangle$ will also decay to
other states or suffers dephasing during the evolution. Moreover,
in Fig.~\ref{fig5}(d), the dephasing rate $\gamma$ is identical to
that in Fig.~\ref{fig5}(c), whereas there exists a slight
deviation of the population with respect to the ideal initial
state ($\epsilon=0.05$). Unfortunately, Fig.~\ref{fig5}(d) also
shows a invalid population transfer. This means that the
traditional STIRAP process is very sensitive to the initial
conditions and the deviation of the initial state produces a
negative  effect. In short, the performances of the traditional
STIRAP process are poor with the pump and Stokes pulses which have
been shown in Fig.~\ref{fig4} without the detuning $\Delta(t)$.

\subsection{Population engineering}

\begin{figure}[htb]
\centering\scalebox{0.35}{\includegraphics{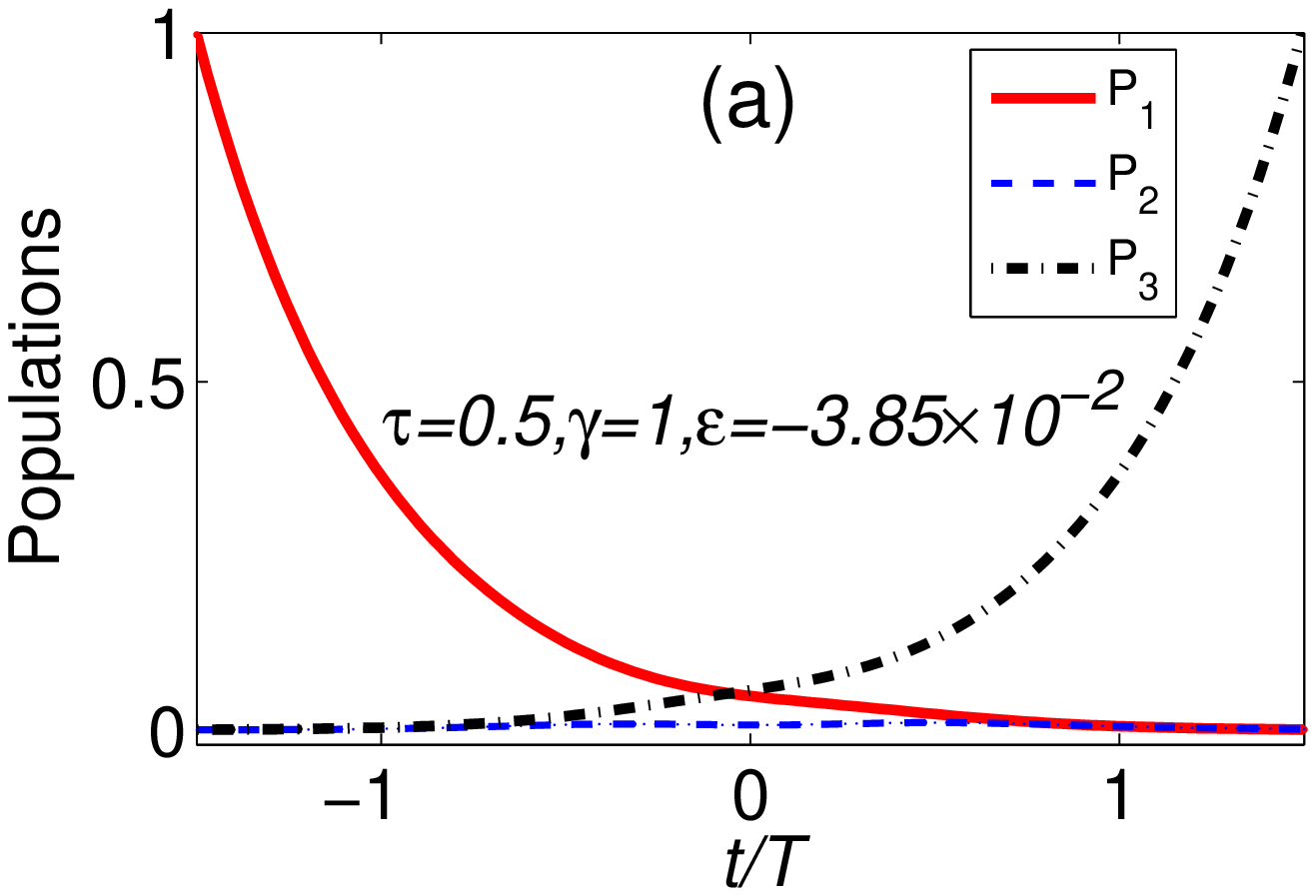}\includegraphics{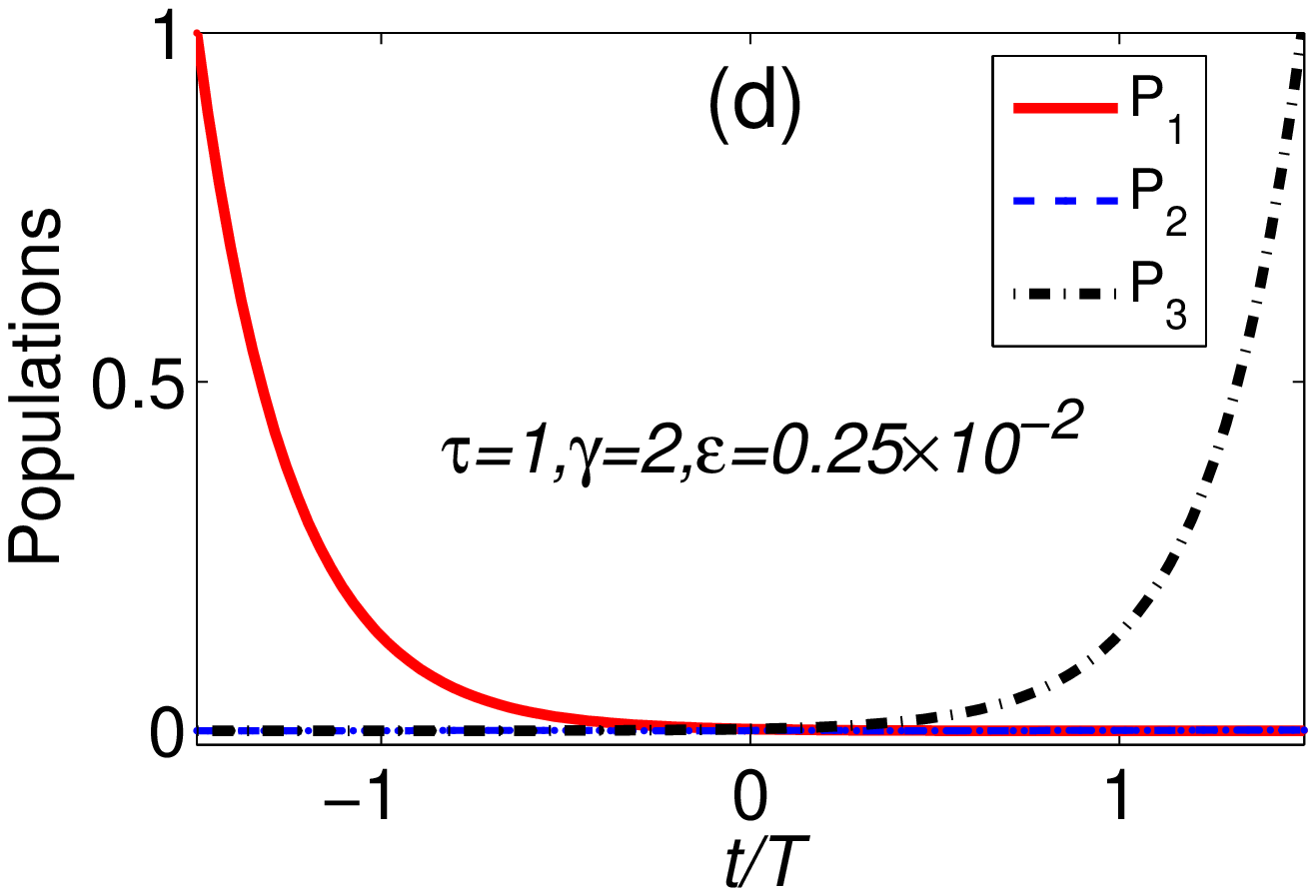}}
\scalebox{0.35}{\includegraphics{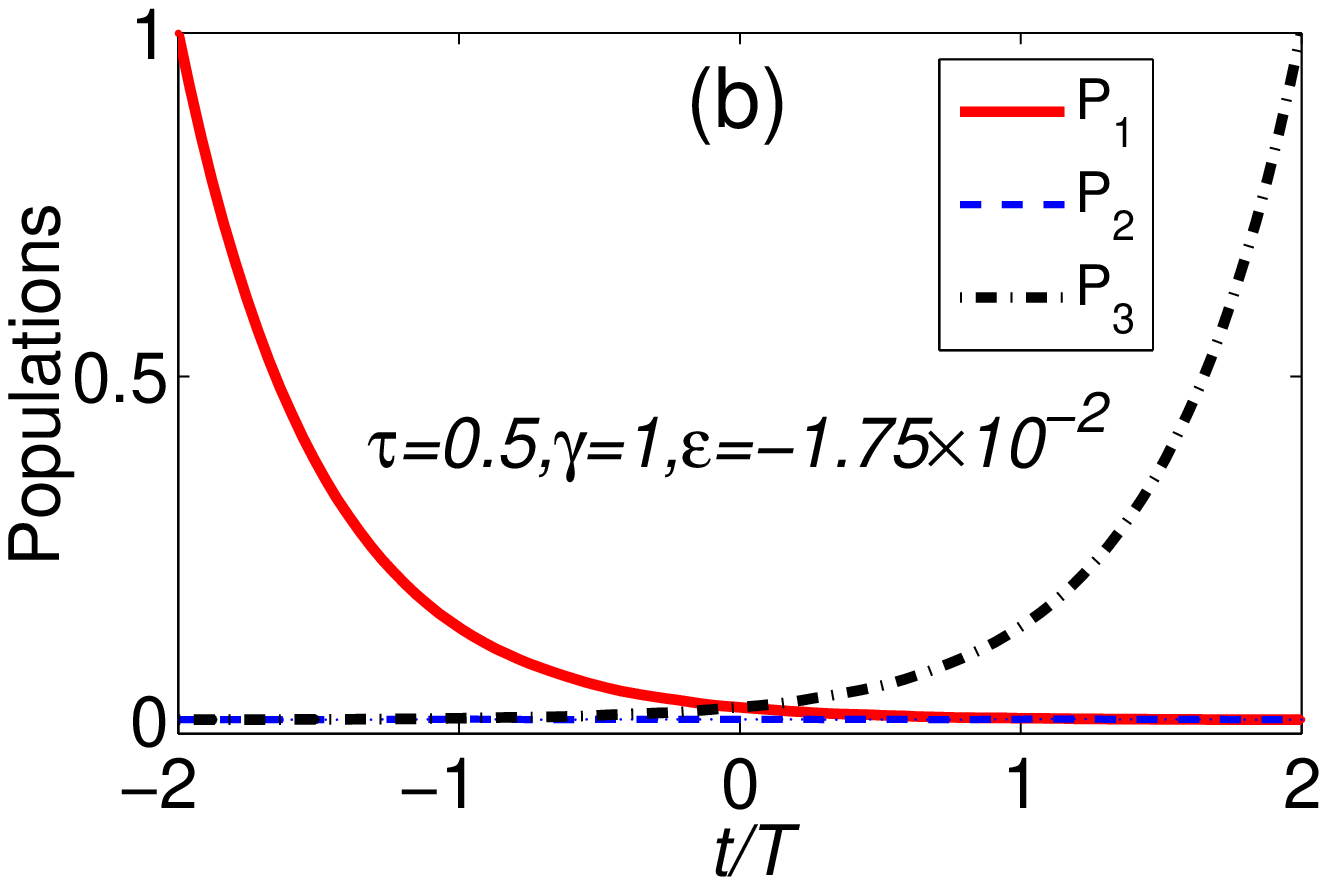}\includegraphics{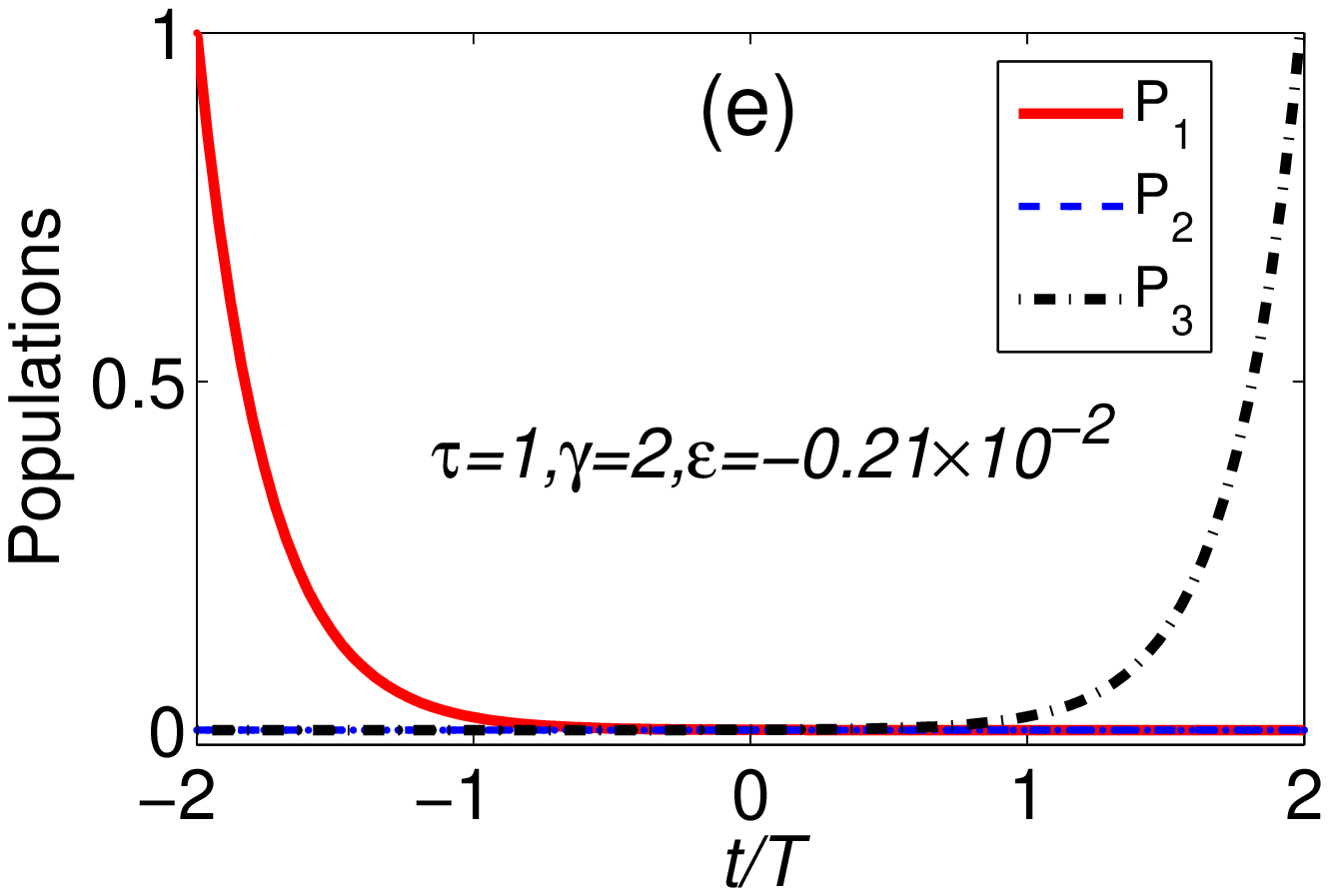}}
\scalebox{0.35}{\includegraphics{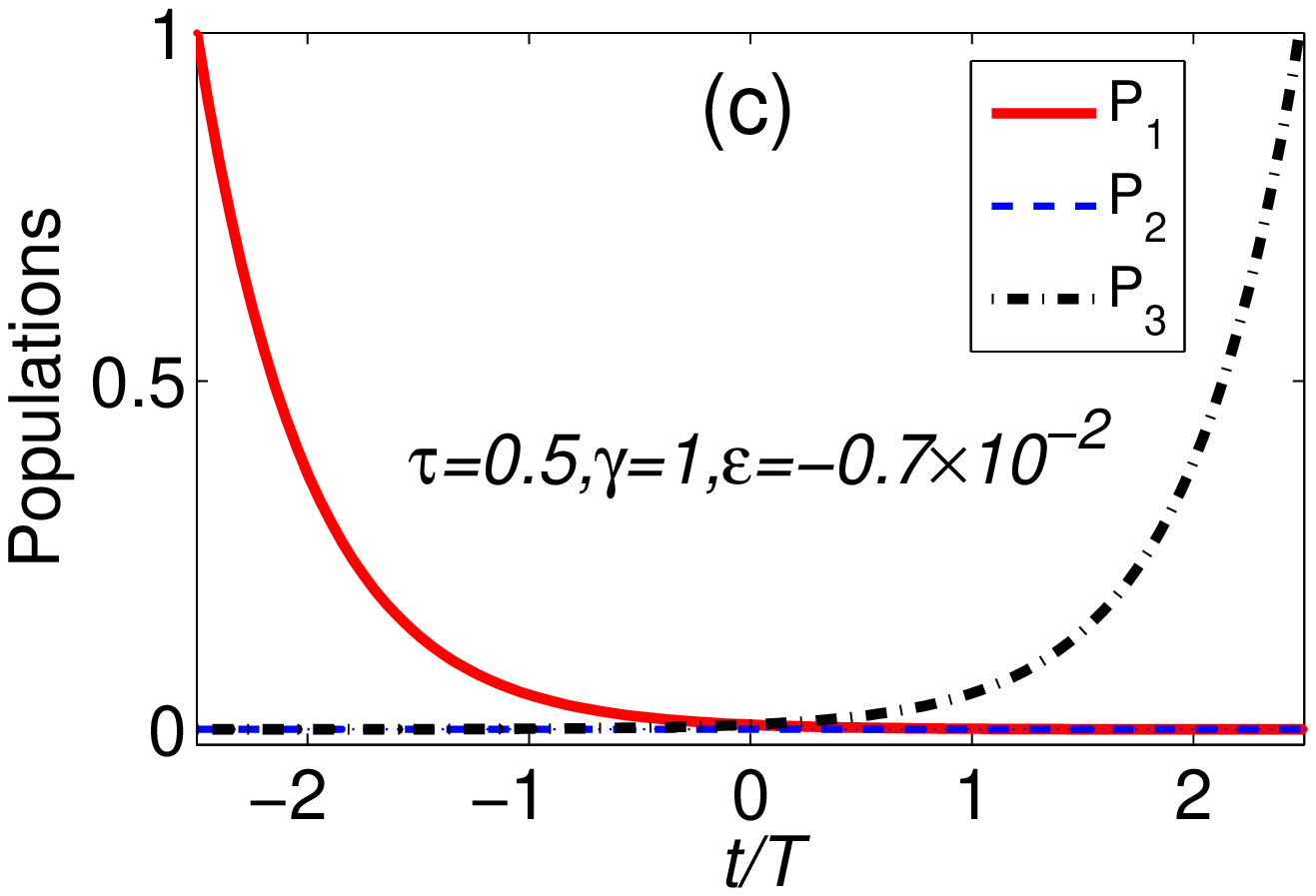}\includegraphics{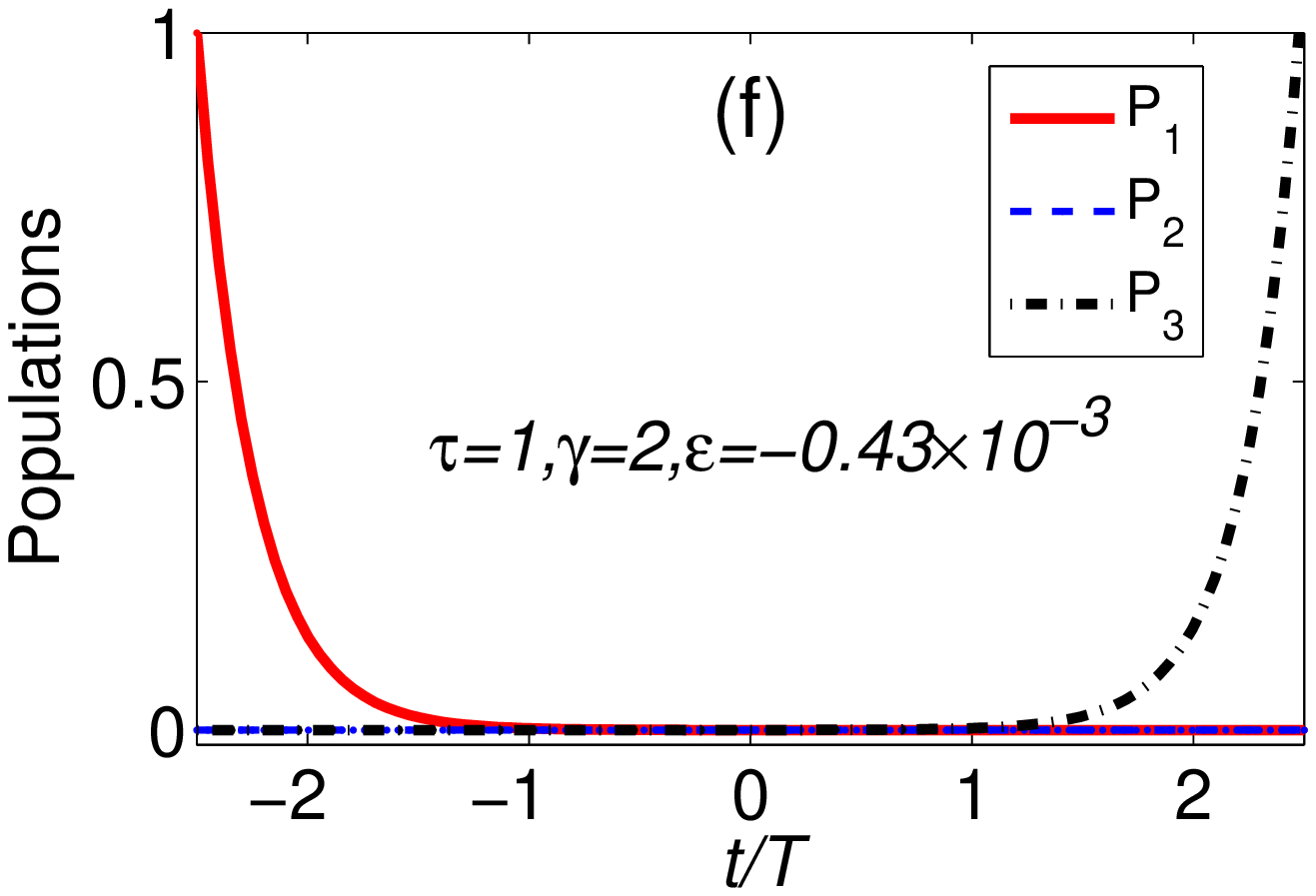}}
\caption{\label{fig6} The time evolution of population $P_1$ ($P_2$,
$P_3$) for the bare state $|1\rangle$ ($|2\rangle$, $|3\rangle$)
with the pump and Stokes pulses and the detuning which have been
shown in Fig.~\ref{fig4} with different imperfect initial states.}
\end{figure}

Now, we start to consider the validity of the theoretical analysis
by numerical calculation  in the following. First, we will study
the population engineering of the system with different
dissipative factors.  Figure~\ref{fig6} shows the time evolution
of populations for the bare states $|1\rangle$, $|2\rangle$,
$|3\rangle$ with the pump and Stokes pulses and the detuning which
have been shown in Fig.~\ref{fig4} for different imperfect initial
states. We can easily find that we can fast achieve a complete
population transfer from $|1\rangle$ to $|3\rangle$ with the
designed pulses and detuning even the initial state is imperfect
($\epsilon\neq0$). More interestingly, the populations of the
state $|2\rangle$ remains negligible all the time in all cases, so
the decay of $|2\rangle$ almost has no effect on the evolution of
the system.  This also means that the evolution of the whole
system is completely governed by the adiabatic state
$|a_0(t)\rangle$ even the initial state is imperfect. In addition,
contrast Fig.~\ref{fig6}(a) with Figs.~\ref{fig6}(b)-(c), we can
find that when the dephasing rate $\gamma$ is fixed, we can
realize the complete population transfer within a shorter time
interval for a relatively large deviation $|\epsilon|$. Similar
results also can be found by contrasting Fig.~\ref{fig6}(d) with
Figs.~\ref{fig6}(e)-(f). What's more, contrast Fig.~\ref{fig6}(a)
with Fig.~\ref{fig6}(d) (Fig.~\ref{fig6}(b) with
Fig.~\ref{fig6}(e) or Fig.~\ref{fig6}(c) with Fig.~\ref{fig6}(f)),
we can find that the $|\epsilon|$ will also signally decrease with
the increase of $\gamma$ with the same STIRAP time interval. The
reason for these results is that the dissipative factors
$\epsilon$ and $\gamma$ are no longer undesirable, actually, they
are the important resources to the scheme. In fact, in principle,
we can realize the complete population transfer from $|1\rangle$
to $|3\rangle$ within an arbitrarily short time if the dissipative
factors are large enough, which is the essential distinction of
the scheme comparing with the previous STIRAP or the shortcut
technique schemes. However, it should be noted here that, since
the system is dissipative, the normalization of the state vector
does not need to be
conserved~\cite{Berry2011,Lavdjpa2012,Chenpra2012} and it will
change in time as shown in Fig.~\ref{fig6}.

In any case the normalized states are popular, since they are
physically very relevant and can be applied directly in the
quantum information processing.  It is advisable to keep that the
norms of the initial and final state vectors are unity. As a
consequence, we will explore in the rest subsection to find an
appropriate approach that can remain normalization of the final
state. Let us now take a closer look at the normalized final state
$|3\rangle$ which is the special form of $|a_{0} ({t_f})\rangle$.
This means that if the amplitude of $|a_{0}({t_f})\rangle$ is 1,
we can obtain the normalized final state $|3\rangle$ naturally. A
naive choice is seeking for the appropriate parameters to enforce
the normalizing, taking into account Eq.~(\ref{eq0-10}) and
Eqs.~(\ref{eq1-2})-(\ref{eq1-3}). Mathematically, we can get some
values for the system parameters (e.g., $\Omega_{s}, \Omega_{p},
\Delta(t), \Gamma(t), \epsilon )$ with such approach. However, the
obvious limitation of this approach is that the solved values
generally are scrambled and might be physically irrelevant in
practice. In addition, taking no account of the difficulty of
calculation, the physical mechanisms behind the parameters are
enshrouded in mist, which remains a obstacle to understanding the
nature of the evolution.

\begin{figure}[htb]
\centering\scalebox{0.35}{\includegraphics{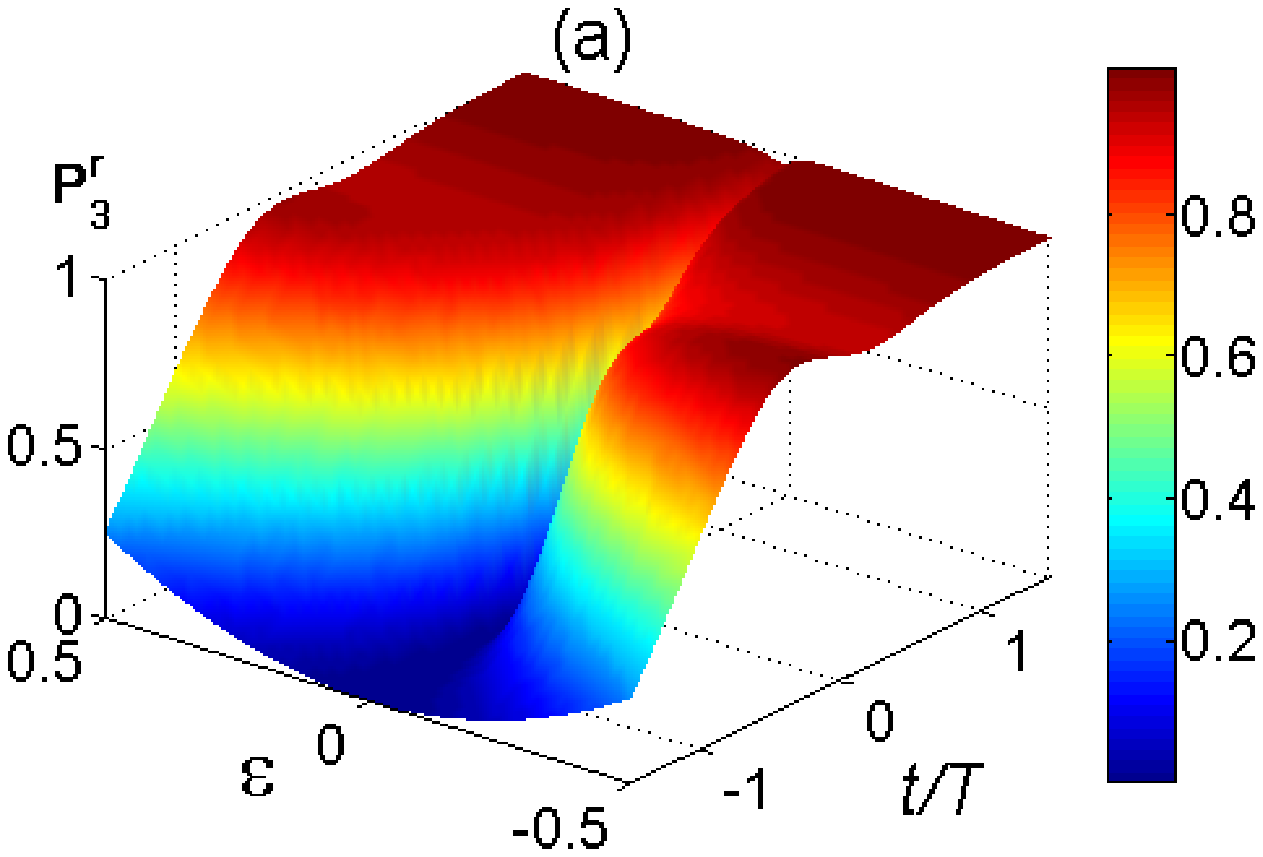}\includegraphics{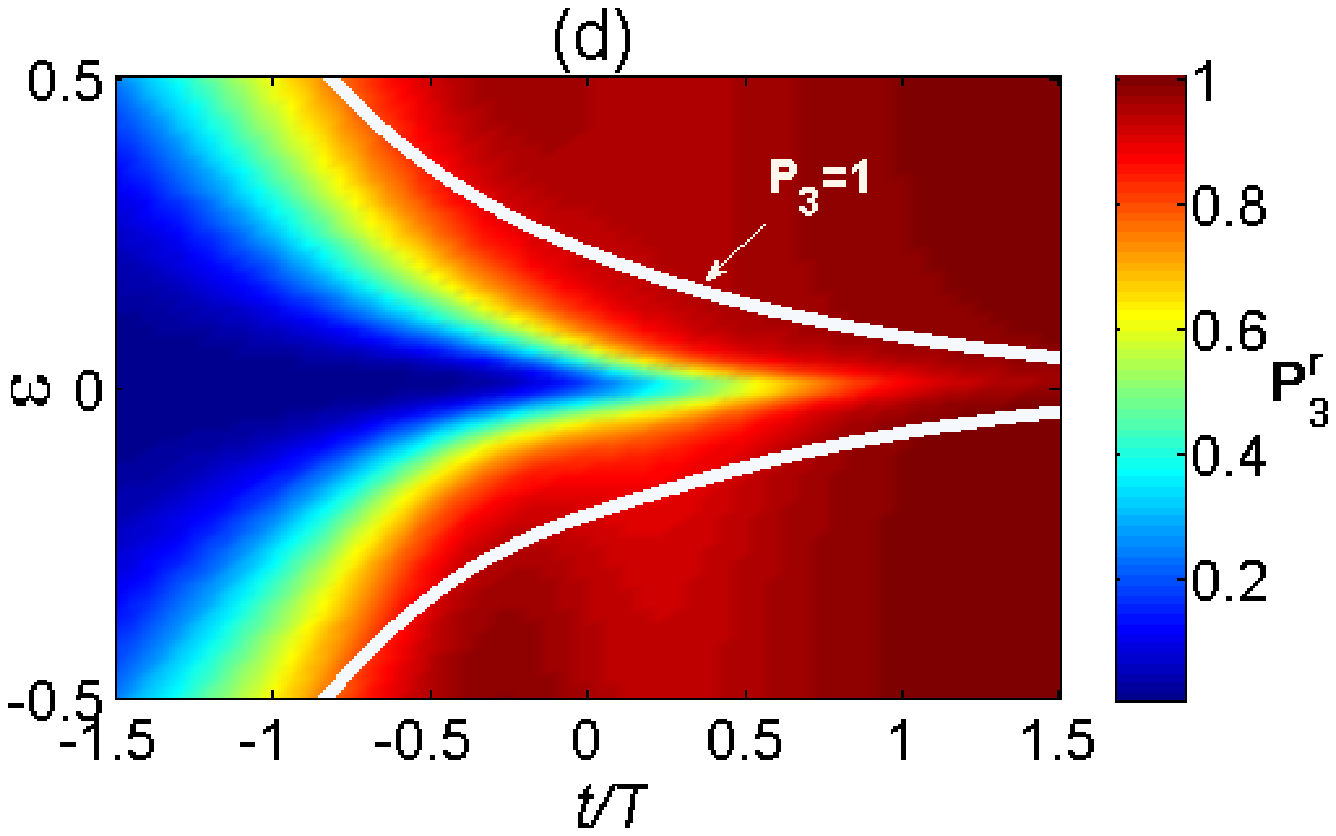}}
\scalebox{0.35}{\includegraphics{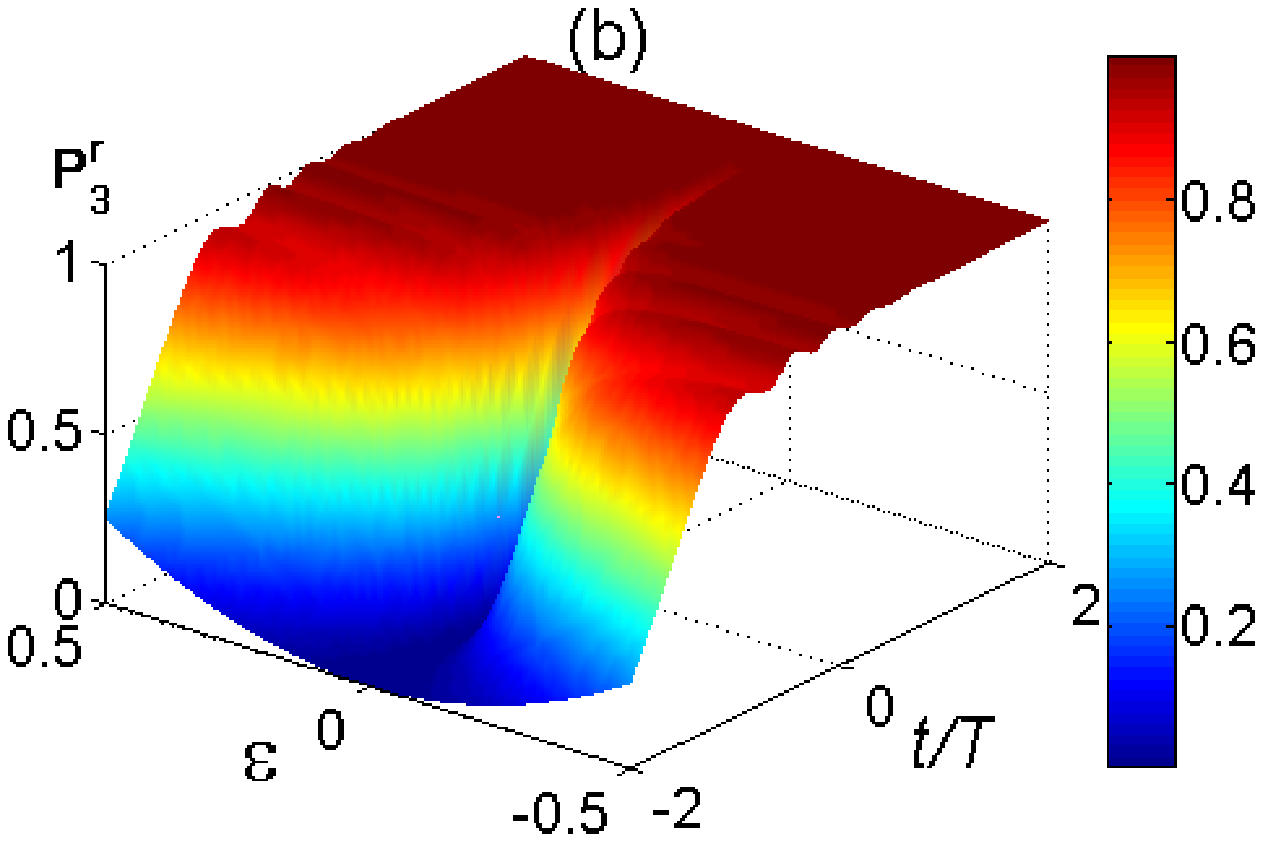}\includegraphics{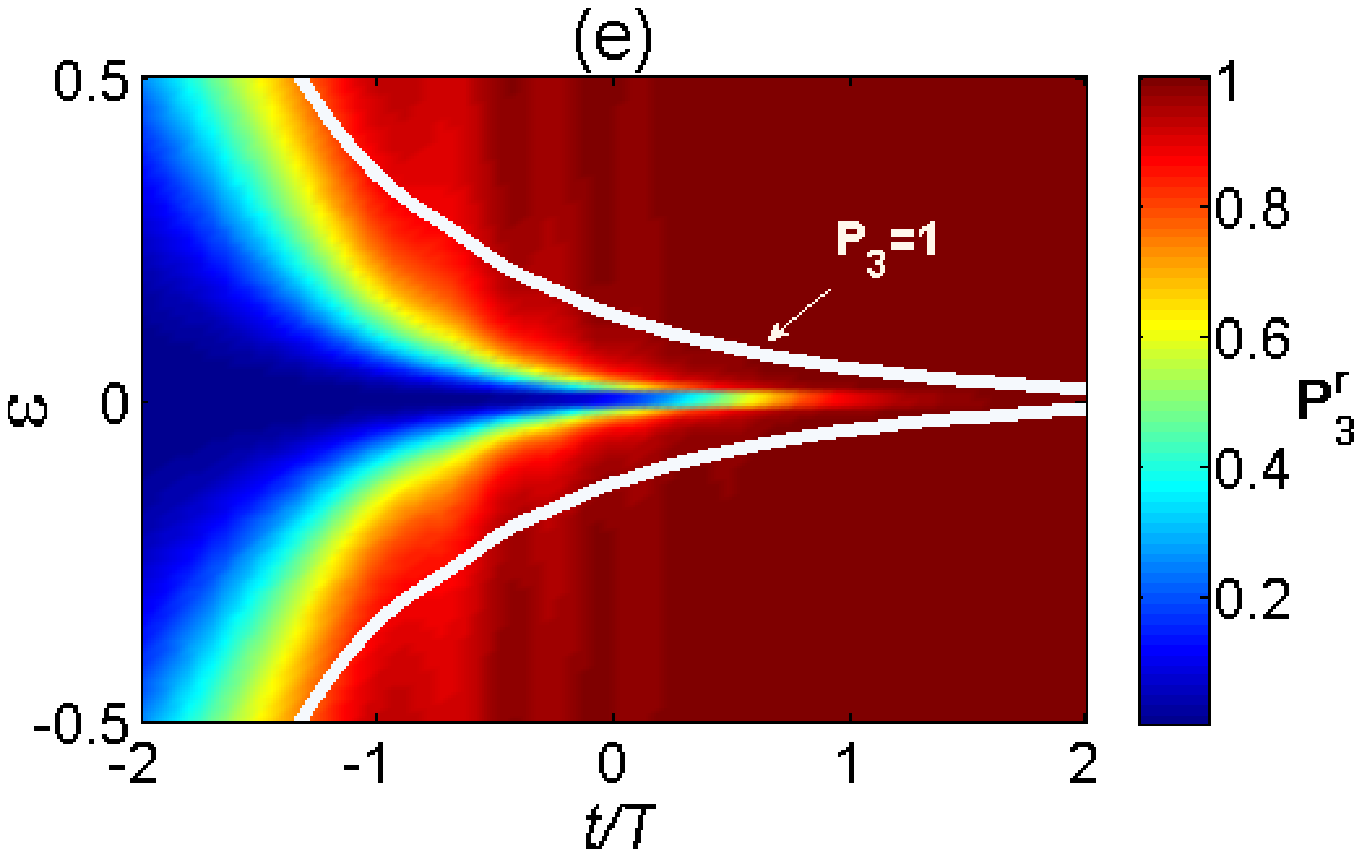}}
\scalebox{0.35}{\includegraphics{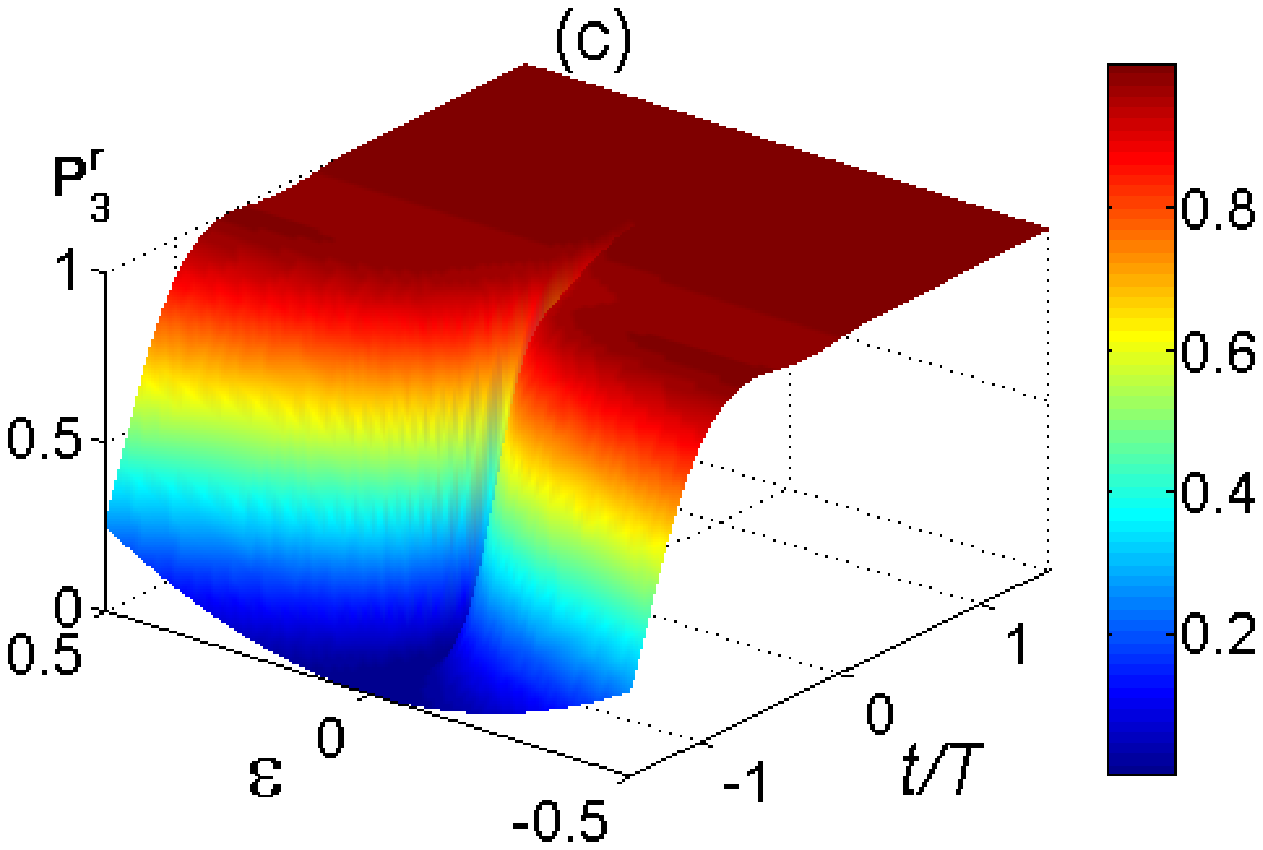}\includegraphics{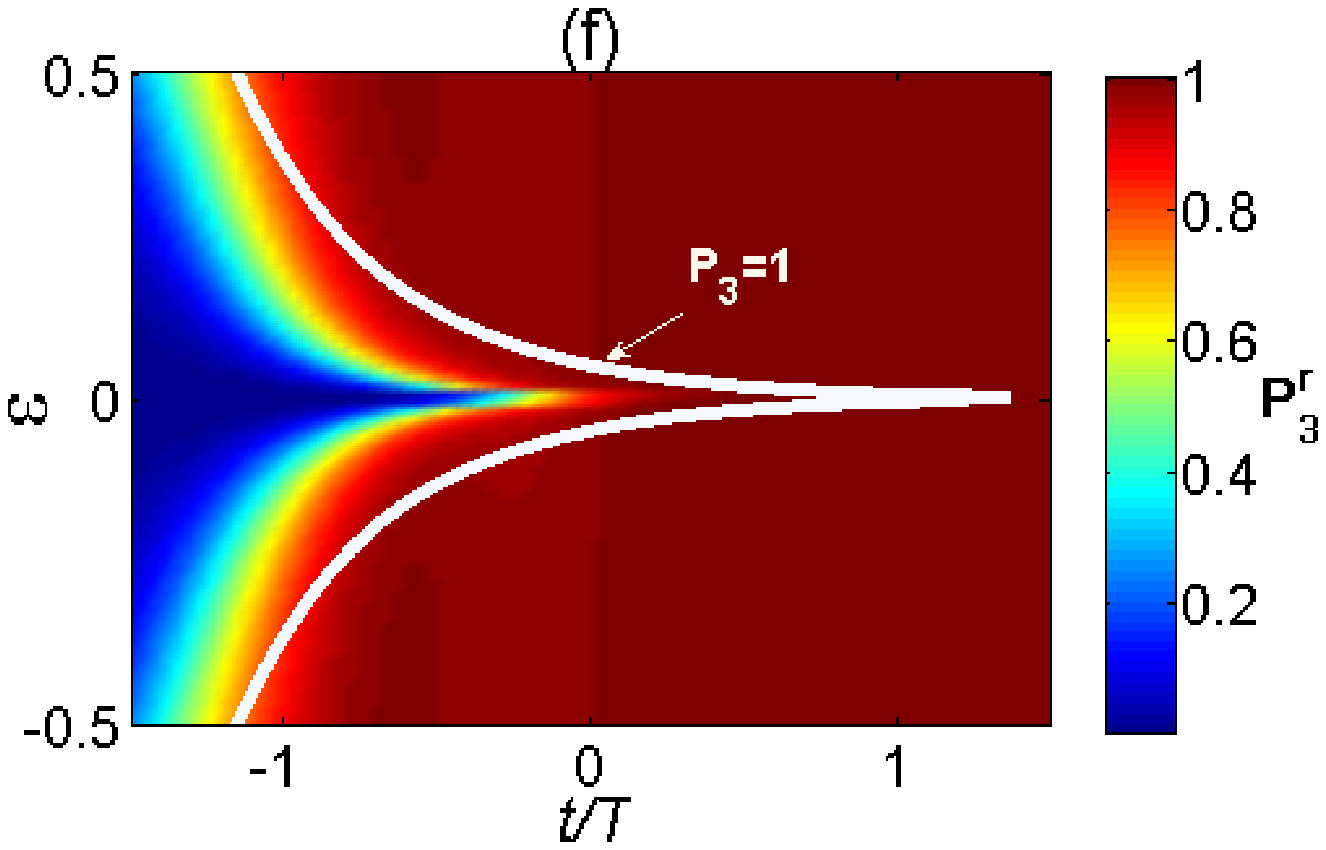}}
\caption{\label{fig7}The relative population $P_3^r$ as a function
of $t$ and $\epsilon$ with three sets of parameters. The
parameters in (a) and (d), (b) and (e), (c) and (f) are identical,
corresponding to $(\gamma=1, t_f=1.5T)$, $(\gamma=1, t_f=2T)$,
$(\gamma=2, t_f=1.5T)$, respectively. The relative population
$P_3^r$ in (a), (b) and (c) with 3D surface, while in (d), (e) and
(f) with two-dimensional (2D) plot.}
\end{figure}

A different approach for imposing the normalization of the final
state is needed. For convenient discussion, we introduce the
concept of relative population $P_i^r~(i=1,2,3)$, where the
relative population is defined as $P_i^r$= $P_i/(P_1+P_2+P_3)$. In
above theoretical analysis, we confirm that there always exists
some optimal parameters to ensure the final state is normalized,
e.g., the parameters in Fig.~\ref{fig6}. Moreover, as shown in
Fig.~\ref{fig6}, in the scheme the population in the state
$|2\rangle$ is negligible and the evolution of the whole system is
completely governed by the adiabatic state $|a_0(t)\rangle$ which
only contains  states $|1\rangle$ and $|3\rangle$. In other words,
so long as we can achieve a complete population transfer within
the evolution time, the relative population $P_3^r$ will reach 1,
More specifically, due to the inherent role of dissipation in the
scheme, the relative population $P_3^r$  will abruptly climb to a
high value, and then change slowly to 1 when the system state is
approaching to the normalized $|3\rangle$. As a consequence,
finding the optimal parameters for imposing the normalization of
the final state is equivalent to the solution to a much simpler
problem, namely, finding the optimal parameters range which will
induce the saltation for the relative population $P_3^r$.

For an intuitive grasp of the approach to normalize the final
state, we plot the time evolution of the relative population for
the bare state $|3\rangle$  with the deviation coefficient
$\epsilon$ in the initial state. As shown in
Figs.~\ref{fig7}(a)-(c), we can get a high relative population
$P_3^r$ for a  wide range of parameters in the red area. In a
sense, if the complete population transfer and robustness  are the
primary concerns, the scheme is quite efficient, since the
relative population $P_3^r$ remains very high (above 0.99) even
releasing the requirements. Furthermore, contrast
Fig.~\ref{fig7}(a) with Figs.~\ref{fig7}(b)-(c), we can find the
range of parameters for reaching a high relative population
$P_3^r$ will signally expand with the increase of $\gamma$ and the
interaction time. Nevertheless, if the normalization of the final
state is also important, the optimum parameters for a complete
population transfer  can be found in the white lines as shown in
Figs.~\ref{fig7}(d)-(f). It is clear from those figures that the
white lines are positioned at the area where the relative
population $P_3^r$ increases very quickly and then changes slowly
to 1 , which is consistent with our deduction. So, we can get the
optimum parameters relations between the deviation coefficient
$\epsilon$ and the evolution time. One can chose the appropriate
time to shut down the designed pluses and detuning to obtain the
normalized final state $|3\rangle$ according to different
$\epsilon$ which is dependent on the practical experiment
conditions. Interestingly, for a relative large $\epsilon$ and
$\gamma$, the population transfer from $|1\rangle$ to $|3\rangle$
can be obtained in a relative shorter time. For example, in
Fig.~\ref{fig7}(f) when $\epsilon$=0.2, we can realize the
complete population transfer from $|1\rangle$ to $|3\rangle$ and
obtain the normalized final state $|3\rangle$ when $t$=-0.745$T$.
In principle, we can realize the complete population transfer
within an arbitrarily short time if the dissipative factors are
large enough, which is the essential distinction of the scheme
comparing with the previous STIRAP or the shortcut technique
schemes.

\subsection{Robustness}

\begin{figure}[htb]
\centering\scalebox{0.4}{\includegraphics{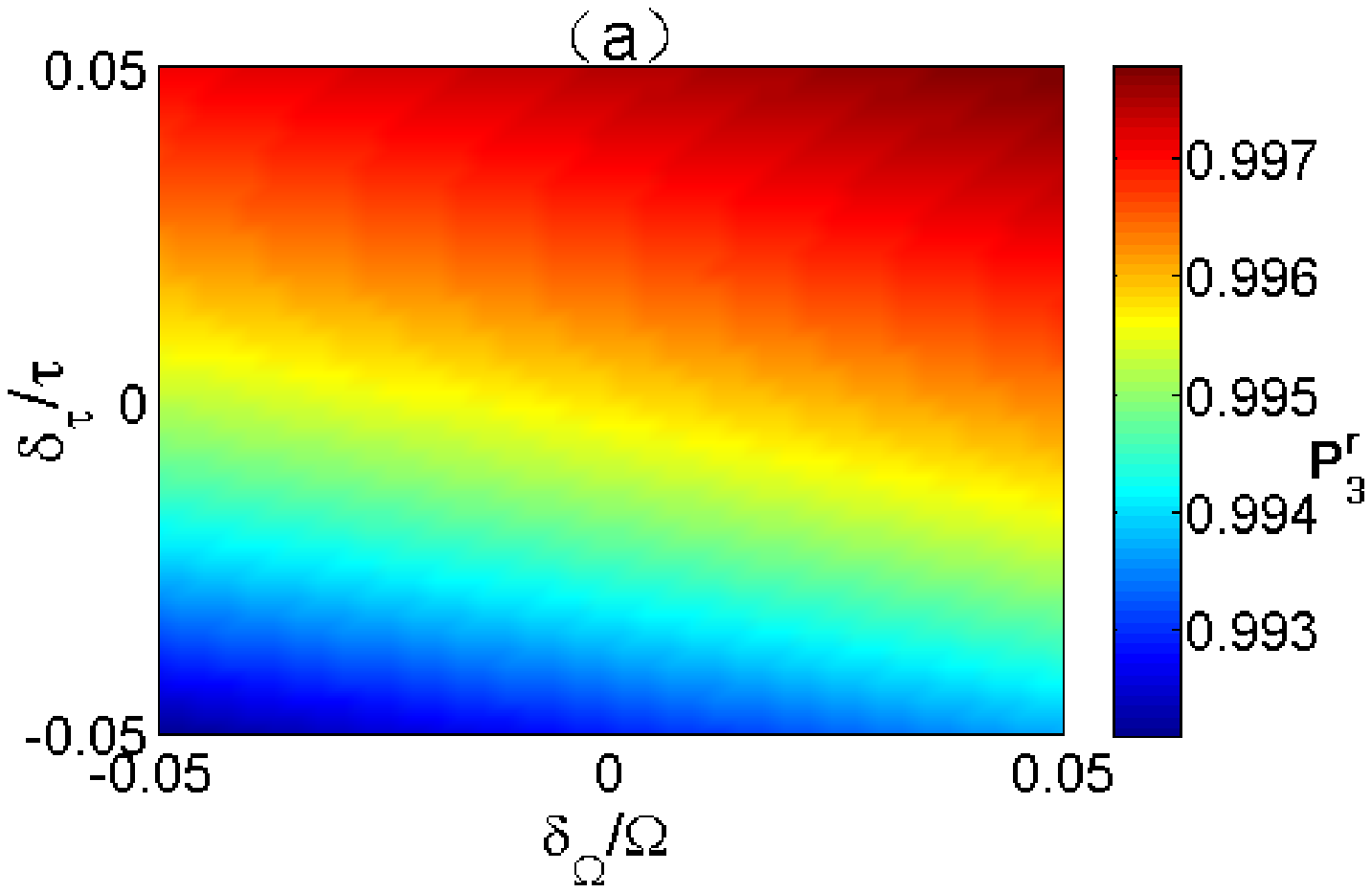}\includegraphics{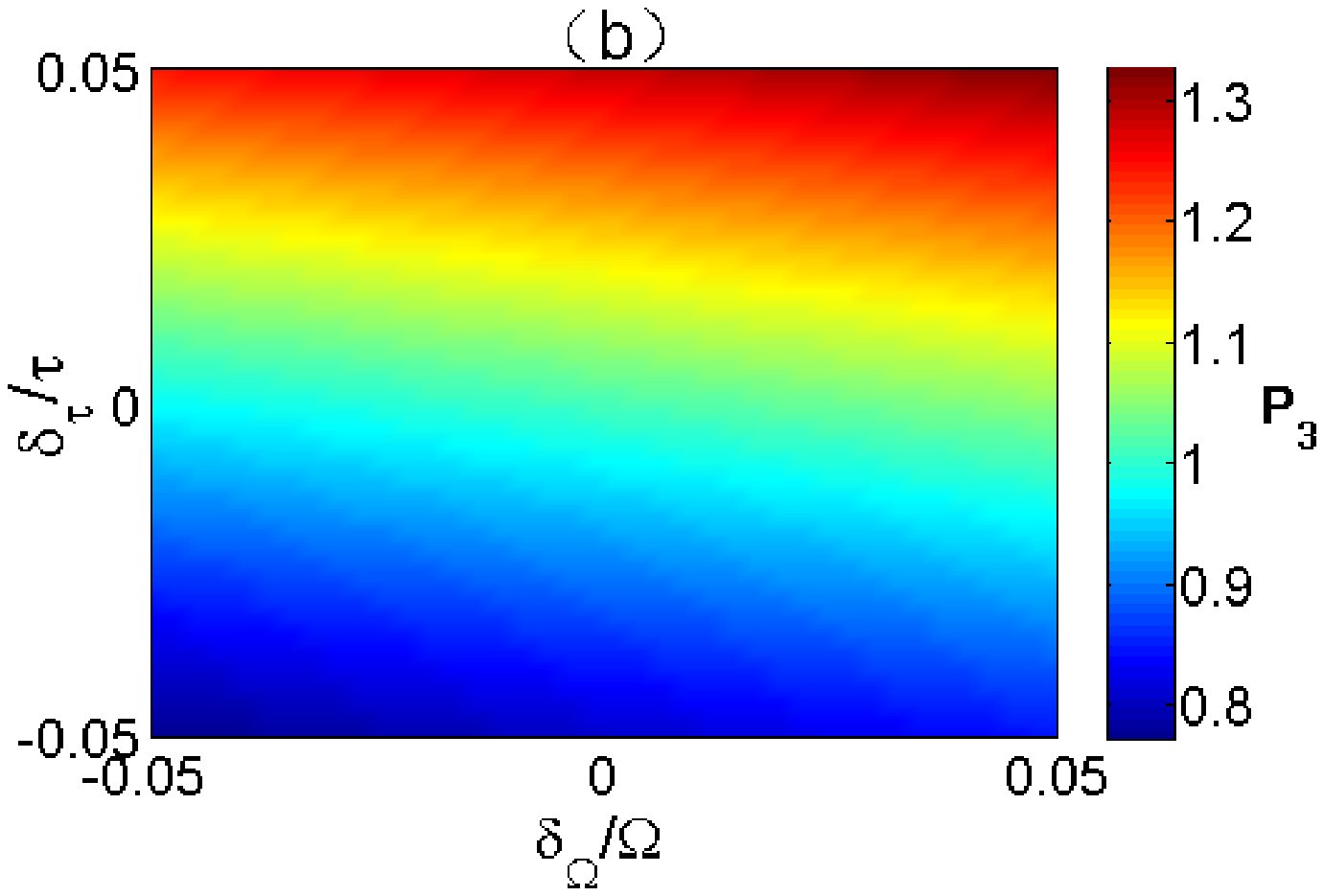}}
\caption{\label{fig8}The relative population $P_3^r$ and the
population $P_3$ vs two relative deviations in the pluses
$\delta_\Omega/\Omega$ and $\delta_\tau/\tau$. The other
parameters are $\Omega=1, \tau=0.5, \gamma=1, t_f=1.5T$,
$\varepsilon=-0.038$, and the detuning with the form of $\Delta_a$
as shown in Table I.}
\end{figure}

In order to test the sensitivity of the improved STIRAP scheme to
the (simulated) variations in the control parameters, we vary the
system parameters around the optimum value for reaching a perfect
population transfer. First, we consider the fluctuatings of the
pluses parameters. In Fig.~\ref{fig8}, the relative population
$P_3^r$ and the population $P_3$ are plotted varying both the peak
Rabi frequencies of pulses and the delay between pulses.
Figure~\ref{fig8}(a) shows that the relative population $P_3^r$ is
insensitive to the fluctuations of the pluses, it only varies in
the $10^{-3}$ order of magnitude with the variation of the pluses
parameters, which indicates the final state almost is populating
on the bare state $|3\rangle$. Nevertheless, the amplitude of
$|3\rangle$ is a little sensitive to the fluctuations of the
pluses as shown in Fig.~\ref{fig8}(b). The population $P_3$ will
vary from 0.77 to 1.3, and we still can obtain approximately a
normalized final state $|3\rangle$ for a moderate fluctuations of
the pluses. Moreover, the effect of the relative deviation in the
time delay $\tau$ will be more obvious than the effect of the
relative deviation of the peak Rabi frequency $\Omega$.  This is
due to the fact the time delay $\tau$ is designed carefully as
shown in Eq.~(\ref{eq2-4}), the relative deviation of $\tau$ will
induce some undesirable coupling transitions which may affect the
normalization of final state slightly. However, the ratio of the
Stokes and pumping pluses won't change dramatically in the
presence of the amplitude variation of peak Rabi frequency
$\Omega$ according to Eq.~(\ref{eq2-4}), so it naturally has a
slight impact on $P_3$.
\begin{figure}[htb]
\centering\scalebox{0.4}{\includegraphics{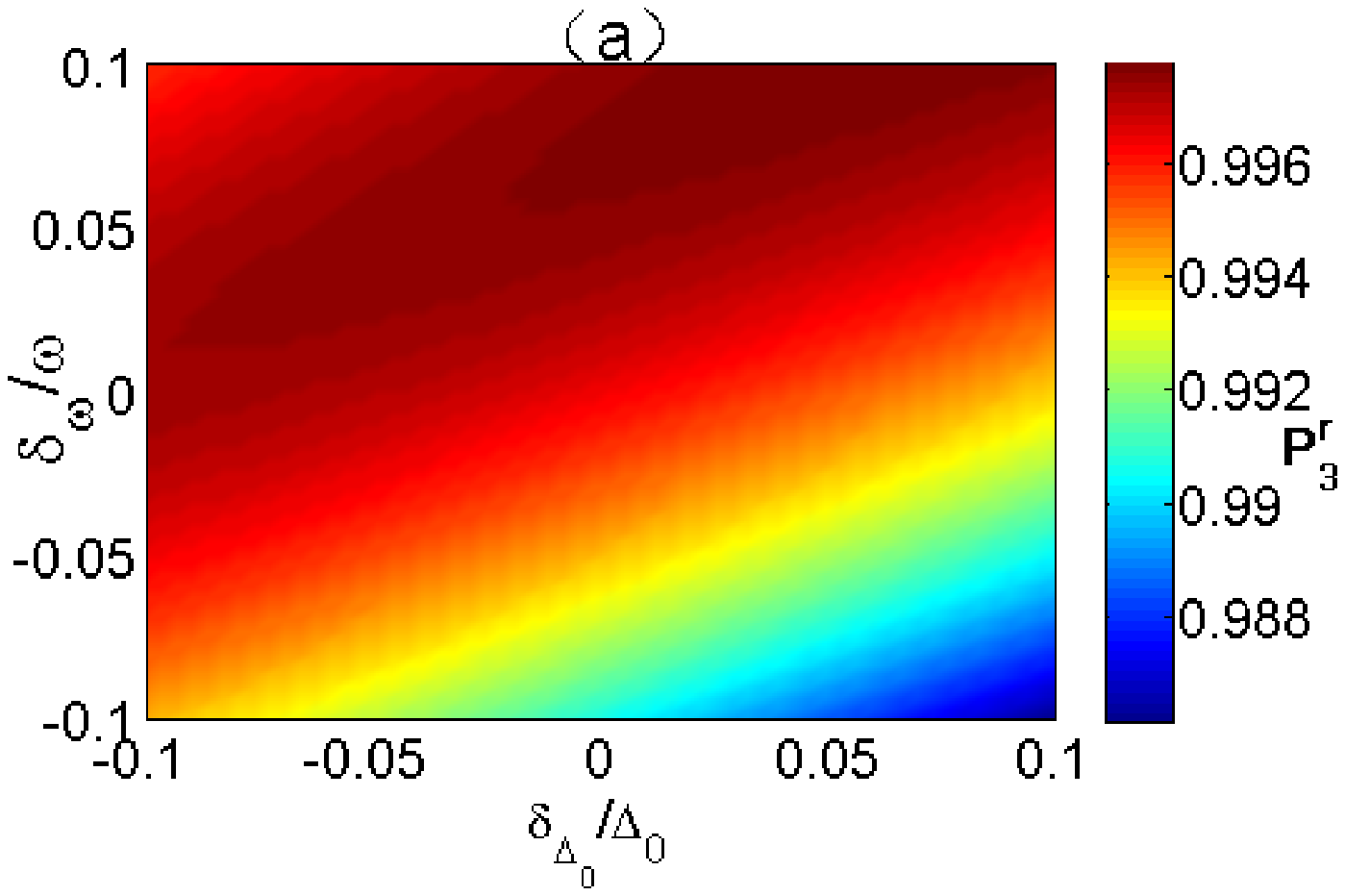}\includegraphics{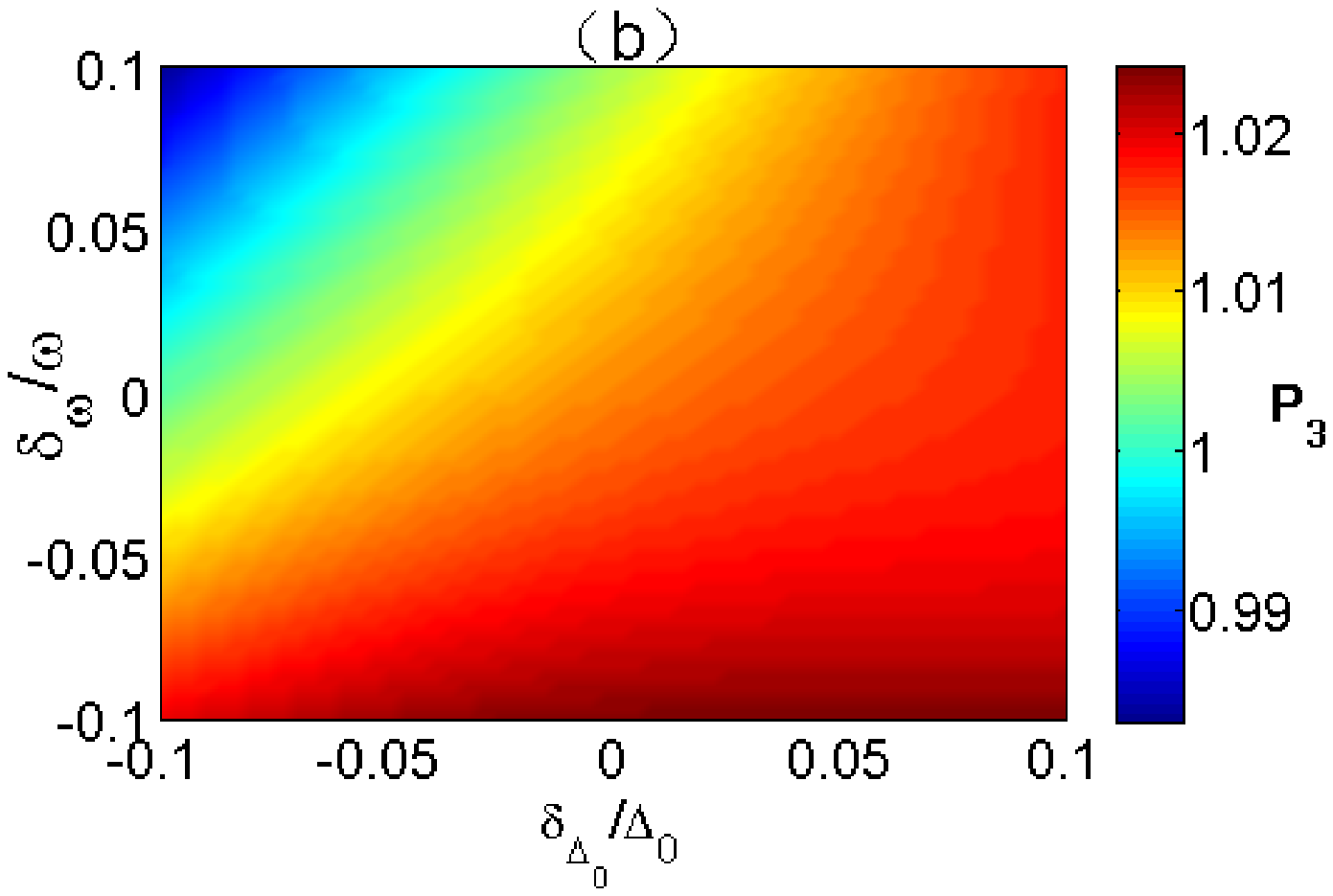}}
\caption{\label{fig9}The relative population $P_3^r$ and the
population $P_3$ vs two relative deviations in the detuning
$\delta_{\Delta_{0}}/{\Delta_{0}}$ and $\delta_\omega/\omega$. The
other parameters are $\Omega=1, \tau=0.5, \gamma=1, t_f=1.5T$,
$\varepsilon=-0.038$, and the detuning with the form of $\Delta_a$
as shown in Table~\ref{table1}.}
\end{figure}
We further test the sensitivity of the scheme to the fluctuations
of the detuning $\Delta_a$ which is introduced in
Table~\ref{table1}. Figure~\ref{fig9} evidences out that both the
relative population $P_3^r$ and population $P_3$ are insensitive
to the fluctuations of the detuning $\Delta_a$, the perfect
$P_3^r$ and population $P_3$ are obtained even if the detuning is
not perfectly matched at the value given in Table~\ref{table1}.

In the above discussion, we calculate the sensitivity of the
improved STIRAP scheme to the variations in the control parameters
by varying one of the protocol parameters in Eq.~(\ref{eq0-2})
(i.e., the pluses parameters and single-photon detuning
parameters) while keeping all other parameters unchanged. It is
nature for us to ask what will happen if all the control
parameters vary simultaneously. Table~2 shows some samples of
relative population $P_3^r$ and the population $P_3$ with
$\delta_\Omega/\Omega$,  $\delta_\tau/\tau$,
$\delta_{\Delta_{0}}/{\Delta_{0}}$, and $\delta_\omega/\omega$. We
can find the relative population $P_3^r$ is insensitive to the
fluctuations of all the control parameters. Nevertheless, the
amplitude of $|3\rangle$ is a little sensitive to the fluctuations
of the control parameters, and it will vary from 0.7774 to 1.3311.
It should be noted that we can approximately obtain the normalized
final state $|3\rangle$, if $|\delta_\tau/\tau|$ is small and the
other control parameters vary moderately. As discussed above, the
scheme shows a good robust feature for potential applications in
quantum manipulation.

\begin{center}
{ {Table 2. Samples of relative population $P_3^r$ and the
population $P_3$ with corresponding $\delta_\Omega/\Omega$,
$\delta_\tau/\tau$, $\delta_{\Delta_{0}}/{\Delta_{0}}$, and
$\delta_\omega/\omega$.}}\\
{\small\begin{tabular}{cccccc}\hline\hline
$\delta_\Omega/\Omega$ \ \ \ \ \ \ &$\delta_\tau/\tau$ \ \ \ \ \ \ \ \ &$\delta_{\Delta_{0}}/{\Delta_{0}}$ \ \ \ \ \ \ \ \ \ &$\delta_\omega/\omega$ \ \ \ \ \ \ \ \ \ &$P_3^r$ \ \ \ \ \ \ \ \ \ \ \ &$P_3$\\
\hline
$5\%\ \ \ \ \ \ \ \ \ \  \ \ $&$5\%\ \ \ \ \ \ \ \ \ \ \ $&$10\%\ \ \ \ \ \ \ \ \ \ \ \ \ $&$10\%\ \ \ \ \ \ \ \ $&$0.9989\ \ \ \ \  \ \ \ \ \ $&$1.3311$\\
$5\%\ \ \ \ \ \ \ \ \ \  \ \ $&$2.5\%\ \ \ \ \ \ \ \ \ \ \ $&$5\%\ \ \ \ \ \ \ \ \ \ \ \ \ $&$5\%\ \ \ \ \ \ \ \ $&$0.9982\ \ \ \ \  \ \ \ \ \ $&$1.1812$\\
$5\%\ \ \ \ \ \ \ \ \ \  \ \ $&$-2.5\%\ \ \ \ \ \ \ \ \ \ \ $&$-5\%\ \ \ \ \ \ \ \ \ \ \ \ \ $&$-5\%\ \ \ \ \ \ \ \ $&$0.9947\ \ \ \ \  \ \ \ \ \ $&$0.9435$\\
$5\%\ \ \ \ \ \ \ \ \ \  \ \ $&$-5\%\ \ \ \ \ \ \ \ \ \ \ $&$-10\%\ \ \ \ \ \ \ \ \ \ \ \ \ $&$-10\%\ \ \ \ \ \ \ \ $&$0.9917\ \ \ \ \  \ \ \ \ \ $&$0.8516$\\
$2.5\%\ \ \ \ \ \ \ \ \ \  \ \ $&$5\%\ \ \ \ \ \ \ \ \ \ \ $&$10\%\ \ \ \ \ \ \ \ \ \ \ \ \ $&$10\%\ \ \ \ \ \ \ \ $&$0.9988\ \ \ \ \  \ \ \ \ \ $&$1.3104$\\
$2.5\%\ \ \ \ \ \ \ \ \ \  \ \ $&$2.5\%\ \ \ \ \ \ \ \ \ \ \ $&$5\%\ \ \ \ \ \ \ \ \ \ \ \ \ $&$5\%\ \ \ \ \ \ \ \ $&$0.9980\ \ \ \ \  \ \ \ \ \ $&$1.1607$\\
$2.5\%\ \ \ \ \ \ \ \ \ \  \ \ $&$-2.5\%\ \ \ \ \ \ \ \ \ \ \ $&$-5\%\ \ \ \ \ \ \ \ \ \ \ \ \ $&$-5\%\ \ \ \ \ \ \ \ $&$0.9944\ \ \ \ \  \ \ \ \ \ $&$0.9239$\\
$2.5\%\ \ \ \ \ \ \ \ \ \  \ \ $&$-5\%\ \ \ \ \ \ \ \ \ \ \ $&$-10\%\ \ \ \ \ \ \ \ \ \ \ \ \ $&$-10\%\ \ \ \ \ \ \ \ $&$0.9913\ \ \ \ \  \ \ \ \ \ $&$0.8322$\\
$-2.5\%\ \ \ \ \ \ \ \ \ \  \ \ $&$5\%\ \ \ \ \ \ \ \ \ \ \ $&$10\%\ \ \ \ \ \ \ \ \ \ \ \ \ $&$10\%\ \ \ \ \ \ \ \ $&$0.9987\ \ \ \ \  \ \ \ \ \ $&$1.2694$\\
$-2.5\%\ \ \ \ \ \ \ \ \ \  \ \ $&$2.5\%\ \ \ \ \ \ \ \ \ \ \ $&$5\%\ \ \ \ \ \ \ \ \ \ \ \ \ $&$5\%\ \ \ \ \ \ \ \ $&$0.9977\ \ \ \ \  \ \ \ \ \ $&$1.1212$\\
$-2.5\%\ \ \ \ \ \ \ \ \ \  \ \ $&$-2.5\%\ \ \ \ \ \ \ \ \ \ \ $&$-5\%\ \ \ \ \ \ \ \ \ \ \ \ \ $&$-5\%\ \ \ \ \ \ \ \ $&$0.9938\ \ \ \ \  \ \ \ \ \ $&$0.8864$\\
$-2.5\%\ \ \ \ \ \ \ \ \ \  \ \ $&$-5\%\ \ \ \ \ \ \ \ \ \ \ $&$-10\%\ \ \ \ \ \ \ \ \ \ \ \ \ $&$-10\%\ \ \ \ \ \ \ \ $&$0.9905\ \ \ \ \  \ \ \ \ \ $&$0.7952$\\
$-5\%\ \ \ \ \ \ \ \ \ \  \ \ $&$5\%\ \ \ \ \ \ \ \ \ \ \ $&$10\%\ \ \ \ \ \ \ \ \ \ \ \ \ $&$10\%\ \ \ \ \ \ \ \ $&$0.9985\ \ \ \ \  \ \ \ \ \ $&$1.2494$\\
$-5\%\ \ \ \ \ \ \ \ \ \  \ \ $&$2.5\%\ \ \ \ \ \ \ \ \ \ \ $&$5\%\ \ \ \ \ \ \ \ \ \ \ \ \ $&$5\%\ \ \ \ \ \ \ \ $&$0.9975\ \ \ \ \  \ \ \ \ \ $&$1.1032$\\
$-5\%\ \ \ \ \ \ \ \ \ \  \ \ $&$-2.5\%\ \ \ \ \ \ \ \ \ \ \ $&$-5\%\ \ \ \ \ \ \ \ \ \ \ \ \ $&$-5\%\ \ \ \ \ \ \ \ $&$0.9934 \ \ \ \  \ \ \ \ \ $&$0.8684$\\
$-5\%\ \ \ \ \ \ \ \ \ \  \ \ $&$-5\%\ \ \ \ \ \ \ \ \ \ \ $&$-10\%\ \ \ \ \ \ \ \ \ \ \ \ \ $&$-10\%\ \ \ \ \ \ \ \ $&$0.9901\ \ \ \ \  \ \ \ \ \ $&$0.7774$\\
\hline \hline
\end{tabular}}
\end{center}

\subsection{The dissipation effects of the excited state}

In the above discussion, we only consider the dissipation effects
of the ground states, the dissipation effects of the excited state
$|2\rangle$ has not been taken into account. However, the latter
is inevitable in practice and may affect the availability of this
method. Thus, we should investigate the influences of the
dissipation effects of the excited state, namely, the damping and
dephasing of the excited state, on the scheme. Furthermore, we
should note that the dissipation effects of the ground states
introduced in the Eq.~(\ref{eq1-0}) is  a good approximation to
the master equation by the quantum trajectory approach. When these
actual decoherence effects are taken into account, the accurate
Lindblad master equation of the whole system can be expressed
as~\cite{Kastoryano2011,Dalibard1992,Plenio1998,D2008}
\begin{eqnarray}\label{3-1}
\dot{\rho}=-i[H_{0}(t),\rho]-\sum_{j=1,2,3}[L_{j}{\rho}L_{j}^{\dag}-\frac{1}{2}(L_{j}^{\dag}L_{j}\rho+{\rho}L_{j}^{\dag}L_{j})],
\end{eqnarray}
where $H_{0}(t)$ is the original Hamiltonian in Eq.~(\ref{eq0-2}),
and $L_{j}$ is the so-called Lindblad operator. In current system,
the damping of the excited state can be described by the Lindblad
operator $L_{1}$, while the dephasing effect can be described by
the Lindblad operator $L_{2}$, in addition, the dissipation
effects of the ground states can be described by the Lindblad
operators $L_{3}$,
\begin{eqnarray}\label{3-2}
L_{1}={\sqrt{\Gamma_{1}}}(|1\rangle\langle{2}|+|3\rangle\langle{2}|),
L_{2}={\sqrt{\Gamma_{2}}}|2\rangle\langle{2}|,
L_{3}={\sqrt{\Gamma}}(|1\rangle\langle{1}|-|3\rangle\langle{3}|),
\end{eqnarray}
where $\Gamma_{1}$ represents the branching ration of the damping
from level $|2\rangle$ to $|1\rangle$ and $|3\rangle$,
$\Gamma_{2}$ denotes the dephasing rate of the $|2\rangle$.

\begin{figure}[htb]
\centering\scalebox{0.4}{\includegraphics{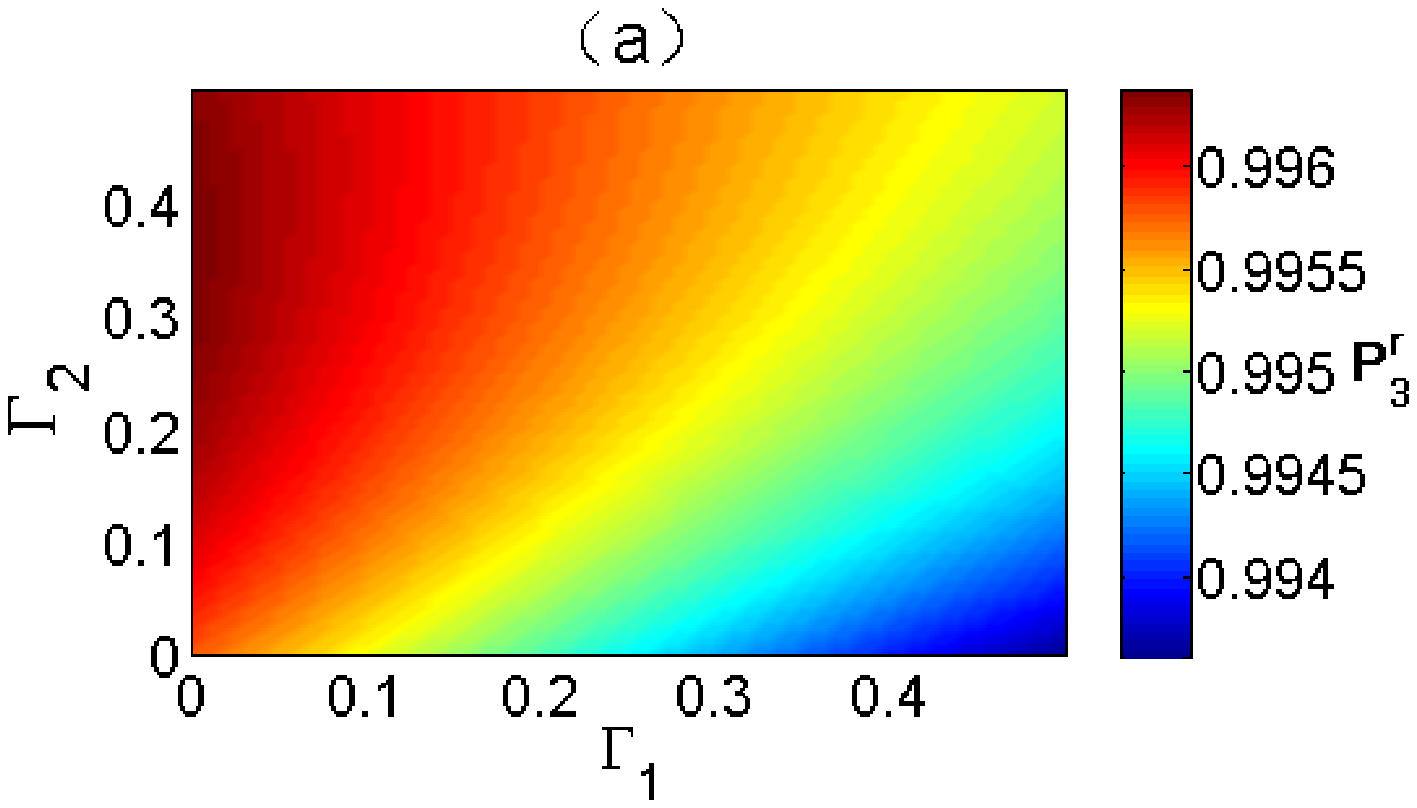}\includegraphics{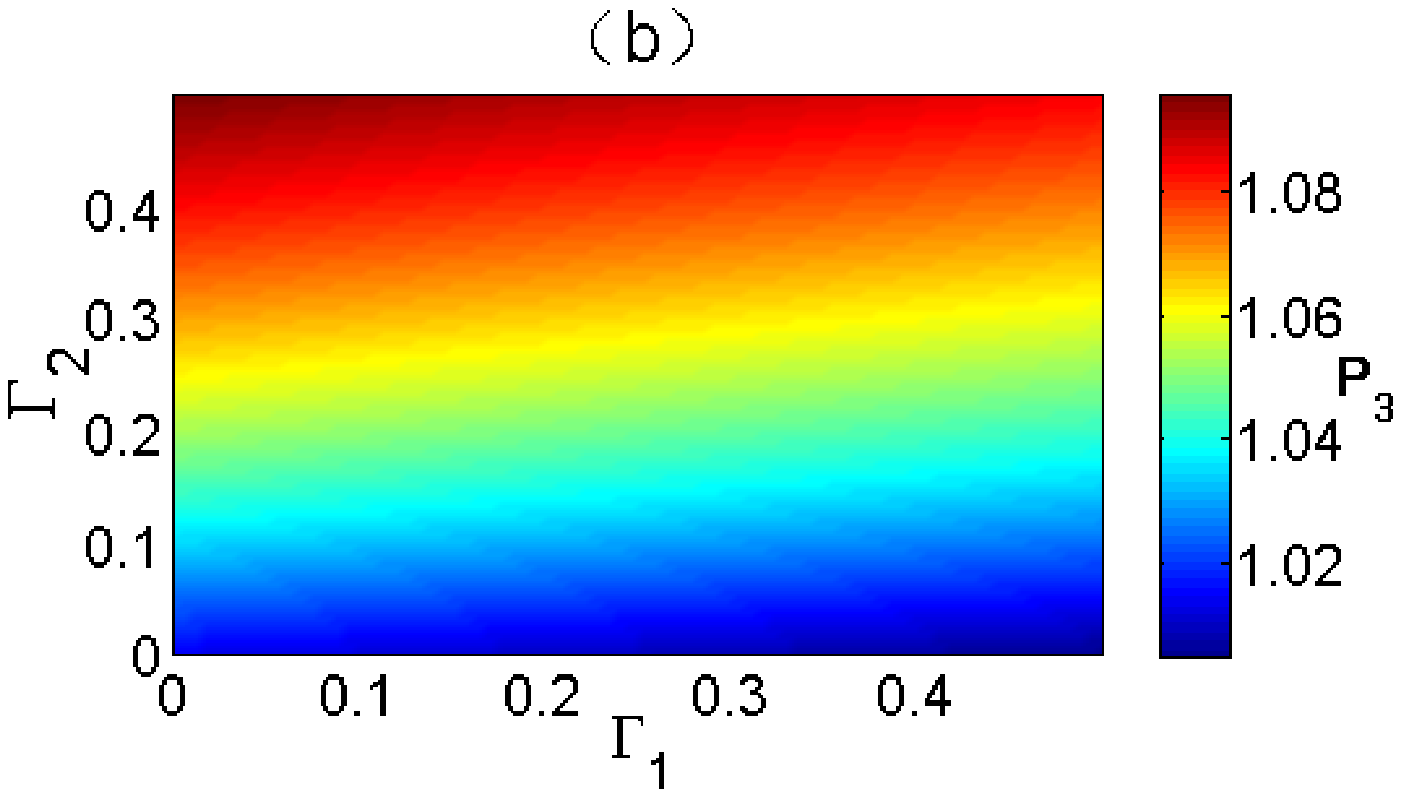}}
\caption{\label{fig10}The relative population $P_3^r$ and the
population $P_3$ vs the decay rates $\Gamma_{1}$ and $\Gamma_{2}$.
The other parameters are the same as shown in the caption of Fig. 6
(a).}
\end{figure}

In this subsection, we will concentrate on the influences of the
decay rates $\Gamma_{1}$ and $\Gamma_{2}$ on the population
transfer, since the influence of $\Gamma$ has been discussed
previously. The relative population $P_3^r$ and the population
$P_3$ vs the decay rates $\Gamma_{1}$ and $\Gamma_{2}$ are given
in Fig.~\ref{fig10}.  It is clear from Fig.~\ref{fig10}(a) that
the relative population $P_3^r$ is insensitive to the dissipation
effects of the excited state, which only varies in the $10^{-3}$
order of magnitude with the variations of $\Gamma_{1}$ and
$\Gamma_{2}$. Meanwhile, Fig.~\ref{fig10}(b) shows that the
normalization of final state $|3\rangle$ will be affected slightly
by the large dissipation effects of the excited state, moreover,
the influence of the dephasing will be more evident than that of
the damping.  This is due to the fact that the excited state is
almost unpopulated during evolution, naturally, the damping rate
of the excited state  has little effect on the system.
Nevertheless, the dephasing will break the superposition of the
states~\cite{Lacour2008} according to the quantum trajectory
theory. As discussed above, the scheme also shows a good robust
feature for potential dissipation effects of the excited state.
Therefore, the scheme can work well in various dissipation cases.

\section{Conclusion}\label{section:V}

We have proposed a scheme to improve the stimulated Raman
adiabatic passage via dissipative quantum dynamics, taking into
account the dephasing effects.  The complete population transfer
can be obtained in a short time with this scheme by the designed
pulses and detuning. We investigate the influences of the
imperfect initial conditions and the dephasing effects in detail.
The numerical simulation results show that the scheme is efficient
and
robust 
since the relative population $P_3^r$ remains very high even
releasing the requirements. Moreover, the dominant dissipative
factors, namely, the dephasing effects of the ground states and
the imperfect initial state are no longer undesirable, on the
contrary, they are the important resources to the scheme.
Interestingly, for a relative large dephasing rate $\gamma$ and
population deviation coefficient $\epsilon$, the population
transfer from $|1\rangle$ to $|3\rangle$ can be obtained in a
relative shorter time. That is, we can realize the complete
population transfer within an arbitrarily short time if the
dissipative factors are large enough. This is the essential
distinction of the scheme comparing with the previous STIRAP or
the shortcut technique schemes. On the other hand, the effects of
variations of the control parameters and various dissipation
effects are also discussed in detail. The results show that the
scheme is insensitive to moderate fluctuations of experimental
parameter and the relatively large dissipation effects of the
excited state, a high fidelity can be reached for a wide range of
parameters. Therefore, the scheme could provide more choices for
the realization of the complete population transfer in the strong
dissipative fields where the standard STIRAP or shortcut schemes
are invalid.

Furthermore, we should note that any quantum system whose
Hamiltonian is possible to be simplified into the form in
Eq.~(\ref{eq1-1}) (the basic for the simplified Hamiltonian can be
arbitrary dressed states, as long as, the dressed states satisfy the
the orthogonality relation and closure relation), the scheme can be
implemented straightforward. This might lead to a useful step toward
realizing fast and noise-resistant quantum information processing
for multi-qubit systems in current technology. Furthermore, one may
also try adding some dissipative factors to make the undesirable
coupling transitions vanish for a more generic system. Moreover, in
above discussion, we consider the damping effect in the Markovian
description for convenience, since some parameters (e.g.,
$\Gamma(t)$) can be considered as the constant. However, in a
general non-Markovian case~\cite{Yupra1999,Jing2010,Maniscalco2006},
these parameters could also depend on time, then, the dynamics of
the system will be complicated and the evolution speed and the
robustness of the scheme will also be changed signally. In fact,
such issue is interesting and attractive, the dynamical aspect of
quantum open systems is also the field where the further extensions
of this work may be explored.

\section*{Acknowledgment}

This work was supported by the National Natural Science Foundation
of China (NSFC) under Grants  No. 11575045 and No. 11374054, and
the Major State Basic Research Development Program of China under
Grant No. 2012CB921601.

\end{document}